\documentclass{article}

\usepackage{microtype}
\usepackage{graphicx}
\usepackage{float}
\usepackage{subcaption}
\usepackage{booktabs}
\usepackage{hyperref}

\usepackage[preprint]{icml2026}

\usepackage{amsmath}
\usepackage{amssymb}
\usepackage{mathtools}
\usepackage{amsthm}

\usepackage{bm}
\usepackage{tabularx}
\usepackage{longtable}
\usepackage{xcolor}

\definecolor{jaxtealdark}{HTML}{16857E}
\definecolor{jaxblue}{HTML}{4285F4}
\definecolor{jaxred}{HTML}{EA4335}
\definecolor{jaxyellow}{HTML}{FBBC05}
\definecolor{jaxgreen}{HTML}{34A853}
\definecolor{jaxcyan}{HTML}{24B6AD}
\definecolor{jaxpurple}{HTML}{9B51E0}

\hypersetup{
    colorlinks=true,
    linkcolor=jaxpurple,
    citecolor=jaxpurple,
    urlcolor=jaxtealdark,
    filecolor=jaxtealdark
}

\usepackage[inline]{enumitem}

\usepackage[capitalize,noabbrev]{cleveref}

\theoremstyle{plain}

\theoremstyle{definition}

\theoremstyle{remark}

\usepackage{array}
\usepackage{amsmath}
\usepackage{amssymb}
\usepackage{mathtools}
\usepackage{amsthm}
\usepackage{bm}

\makeatletter
\newcommand{\dlmf}[1]{%
\citep[%
  \def\nextitem{\def\nextitem{, }}%
  \@for \el:=#1\do{\nextitem\href{http://dlmf.nist.gov/\el}{(\el)}}%
]{olver_nist_2010}%
}
\makeatother

\usepackage[textsize=tiny]{todonotes}

\usepackage[most]{tcolorbox}

\newtcolorbox{promptbox}[1][]{
  enhanced,
  breakable,
  colback=gray!5!white,
  colframe=black!70!white,
  fonttitle=\bfseries\sffamily,
  title=#1,
  boxrule=0.5pt,
  arc=3pt,
  left=8pt, right=8pt, top=8pt, bottom=8pt,
  shadow={2pt}{-2pt}{0mm}{black!15!white} 
}

\icmltitlerunning{\texttt{gyaradax}: Local Gyrokinetics JAX Code}

\begin{document}

\twocolumn[
\icmltitle{
\raisebox{-5.0pt}{\includegraphics[width=1.2em]{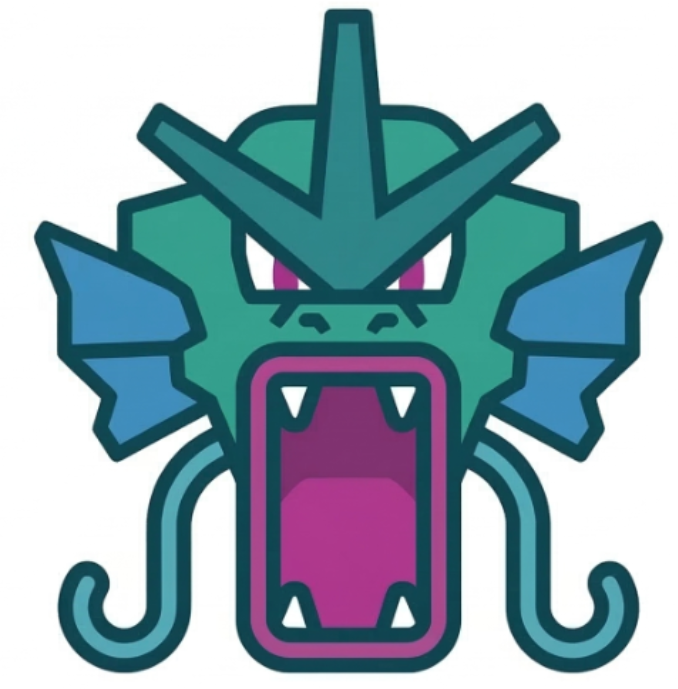}}\hspace{0.2em}%
\texttt{gyaradax}: Local Gyrokinetics JAX Code
}
  \icmlsetsymbol{equal}{*}
  \vspace{-1.2em}
  \begin{icmlauthorlist}
    \icmlauthor{Gianluca Galletti}{equal,jku}
    \icmlauthor{Eric Volkmann}{equal,jku}
    \icmlauthor{Johannes Brandstetter}{jku,emmi}
  \end{icmlauthorlist}

  \icmlaffiliation{jku}{Institute for Machine Learning, JKU Linz,}
  \icmlaffiliation{emmi}{EMMI AI}

  \icmlcorrespondingauthor{Gianluca Galletti}{g.galletti@ml.jku.at}

\begin{center}
    \href{https://github.com/gerkone/gyaradax}{%
        \raisebox{-0.3em}{\includegraphics[width=1.15em]{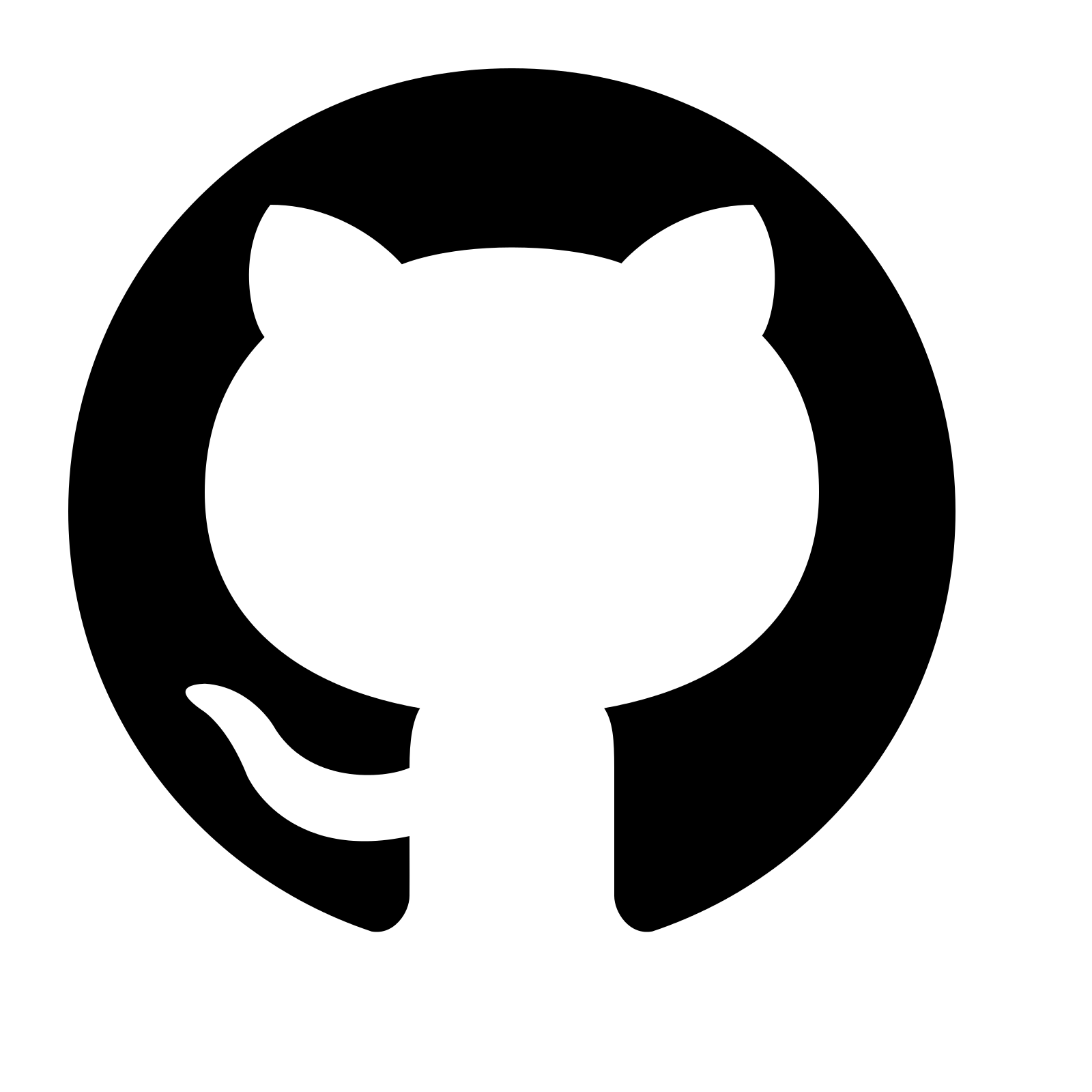}}
        \textbf{\texttt{gerkone/gyaradax}}%
        \vspace{0.5em}
    }
\end{center}

  \icmlkeywords{Gyrokinetics, Differentiable Solvers, JAX, Coding Agents}
]

\makeatletter
\begingroup
\long\def\@makefntext#1{\noindent #1}
\printAffiliationsAndNotice{\par} 
\endgroup
\makeatother

\begin{abstract}
Gyrokinetic simulations are essential for understanding and controlling turbulence in fusion plasmas, yet they are oftentimes implemented in legacy codebases, in many cases CPU-bound. These are both hard to maintain and especially incompatible with optimization and ML workflows.
\texttt{gyaradax} is a minimal JAX/CUDA solver for local flux-tube gyrokinetics. We base our implementation on GKW \cite{PEETERS_GKW_2009}, but with added native GPU acceleration and automatic differentiation. We validate \texttt{gyaradax} against analytical cases and empirical benchmarks, achieving formal agreement and statistical parity with GKW alongside a substantial speedup.
We deliberately and extensively utilized agentic workflows in this project.
A key contribution is showing that coding agents, guided by human expertise, structured prompting, and measurable progress through unit testing enabled extremely fast translation of complex Fortran code, and further optimizations.
\texttt{gyaradax} facilitates research at the intersection of ML and plasma physics. We showcase this through practical examples in inverse problems and sensitivity analysis.
\end{abstract}
\vspace{-8px}

\section{Introduction}

Plasma turbulence is one of the most challenging multiscale problems in computational physics. The community has for decades relied on highly optimized and widely accepted codebases like GKW \cite{PEETERS_GKW_2009} and GENE \cite{jenko2000electron}. Built for CPU clusters, they rely on complex MPI parallelization that makes them quite rigid. Although newer codes like CGYRO \cite{CANDY201673} and modern versions of GENE offer GPU support, their underlying architecture still lacks the flexibility to easily expand or integrate into ML workflows.

In this work, we present \texttt{gyaradax}, a slim JAX code for local flux-tube gyrokinetics, supporting both adiabatic and kinetic electron models.
We pick JAX \cite{jax2018github} following the path of recent works across scientific domains, for example JAXFLUIDS \cite{Bezgin2023} and JAX-SPH \cite{toshev2024jax} for fluid dynamics, TORAX \cite{citrin2024toraxfastdifferentiabletokamak} for fast tokamak transport simulation, and JAX-MD \cite{jaxmd2020} for molecular dynamics.
\texttt{gyaradax} bridges the gap between legacy gyrokinetics and the JAX ecosystem, enabling hardware-accelerated automatic differentiation for plasma turbulence research \cite{paischer2025gyroswin,kelling2025artificialscientistintransit,Zanisi2024}.
A welcome side effect of using JAX is the increased accessibility: the core integrator and field solver are implemented in approximately 3\,000 lines of JAX, compared to over 30\,000 lines of functionally equivalent Fortran.

We developed \texttt{gyaradax} with the heavy use of coding agents \cite{yang2024sweagentagentcomputerinterfacesenable} and \textit{vibecoding} \cite{karpathy2025vibecoding},
applying emerging AI-driven software engineering practices to complex scientific codebases \cite{chen2025scienceagentbenchrigorousassessmentlanguage,duston2025ainsteinbenchbenchmarkingcodingagents}.
A central challenge in deploying agents is oversight \cite{bowman2022measuringprogressscalableoversight}: \textit{how can we verify that generated code is correct as systems grow in complexity?}
For numerical codebases this issue is naturally dampened, as correctness can be rigorously verified against analytical cases and reference results or implementations.
With this strong ``reward'' signal coupled with (aspiring) experts, we find that coding agents can effectively navigate and implement extensive scientific computing frameworks, freeing resources for more fundamental research.

In this work, we outline the underlying gyrokinetic equations (\cref{sec:gyrokinetics}), the implementation and optimization of \texttt{gyaradax} (\cref{sec:gyaradax}), and notes on the agent-assisted development process (\cref{sec:vibecoding}).
We formally verify our solver on the \citet{Rosenbluth} zonal flow test and on the Cyclone Base Case \cite{cbc}. Furthermore, we run extensive empirical evaluations against GKW \cite{PEETERS_GKW_2009} reference data, comparing statistical quantities (time-averaged transport fluxes, wavenumber spectra, growth rates) across a broad set of equilibrium configurations for both adiabatic and kinetic electrons (\cref{sec:verification}).
Finally, we demonstrate differentiable programming experiments, including gradient-based recovery of $R/L_T$ and sensitivity analysis of growth rates (\cref{sec:exp}).


\section{Gyrokinetics}
\label{sec:gyrokinetics}
Within computational plasma physics, gyrokinetics~\citep{frieman1982nonlinear,Krommes_gyrokinetics_2012} is the foundational framework for turbulent transport modeling.
By averaging over the fast gyro-angle dependency, it decouples high-frequency modes, isolating low-frequency fluctuations at the gyroradius scale while preserving finite Larmor radius effects in the distribution function $f$. This effectively reduces the 6D Vlasov-Maxwell equations to a 5D guiding-center formulation, making the system numerically tractable.

To this end, \texttt{gyaradax} uses the standard $\delta f$ form \citep{frieman1982nonlinear}, decomposing $f$ into a stationary Maxwellian background $F_M$ and a fluctuating component $\delta f(v_\parallel, \mu, s, \mathbf{k}_\perp)$.
Here, $v_\parallel$ and $\mu$ are the parallel velocity and magnetic moment, $s$ goes along the magnetic field line, and $\mathbf{k}_\perp$ includes radial $k_\psi$ ($k_x$) and bi-normal $k_\zeta$ ($k_y$).  

\subsection{Physical Components}
We clearly mark each term in the right-hand side of the Vlasov-Poisson equation (I to VIII), and adopt a naming convention similar to \citet{PEETERS_GKW_2009}, 
\begin{equation*}
\resizebox{\columnwidth}{!}{$\displaystyle
\begin{aligned}
    \frac{\partial f}{\partial t} = \hspace{-5px}
    & \underbrace{\colorbox{jaxpurple!15}{$\displaystyle \vphantom{\frac{Z e F_M}{T} \left( v_\parallel \nabla_\parallel \bar{\phi} + i(\mathbf{k}_\perp \cdot \mathbf{v}_D) \bar{\phi} \right)} - v_\parallel \nabla_\parallel f$}}_{\text{Parallel Advection (I)}} 
    \underbrace{\colorbox{jaxpurple!15}{$\displaystyle \vphantom{\frac{Z e F_M}{T} \left( v_\parallel \nabla_\parallel \bar{\phi} + i(\mathbf{k}_\perp \cdot \mathbf{v}_D) \bar{\phi} \right)} - i(\mathbf{k}_\perp \cdot \mathbf{v}_D) f$}}_{\text{Magnetic Drift (II)}}
    \underbrace{\colorbox{jaxpurple!15}{$\displaystyle \vphantom{\frac{Z e F_M}{T} \left( v_\parallel \nabla_\parallel \bar{\phi} + i(\mathbf{k}_\perp \cdot \mathbf{v}_D) \bar{\phi} \right)} + \mu \nabla_\parallel B \frac{\partial f}{\partial v_\parallel}$}}_{\text{Mirror Term (IV)}} \\
    & \underbrace{\colorbox{jaxtealdark!15}{$\displaystyle \vphantom{\frac{Z e F_M}{T} \left( v_\parallel \nabla_\parallel \bar{\phi} + i(\mathbf{k}_\perp \cdot \mathbf{v}_D) \bar{\phi} \right)} - \mathbf{v}_E \cdot \nabla F_M$}}_{\text{Equilibrium Drive (V)}} 
    \underbrace{\colorbox{jaxtealdark!15}{$\displaystyle \vphantom{\frac{Z e F_M}{T} \left( v_\parallel \nabla_\parallel \bar{\phi} + i(\mathbf{k}_\perp \cdot \mathbf{v}_D) \bar{\phi} \right)} - \frac{Z e F_M}{T} \left( v_\parallel \nabla_\parallel \bar{\phi} + i(\mathbf{k}_\perp \cdot \mathbf{v}_D) \bar{\phi} \right) $}}_{\text{Field Drives (VII and VIII)}} \\
    & \underbrace{\colorbox{jaxred!15}{$\displaystyle \vphantom{\frac{Z e F_M}{T} \left( v_\parallel \nabla_\parallel \bar{\phi} + i(\mathbf{k}_\perp \cdot \mathbf{v}_D) \bar{\phi} \right)} - \mathbf{v}_E \cdot \nabla_\perp f$}}_{\text{Nonlinear (III)}}
    \underbrace{\colorbox{jaxblue!15}{$\displaystyle \vphantom{\frac{Z e F_M}{T} \left( v_\parallel \nabla_\parallel \bar{\phi} + i(\mathbf{k}_\perp \cdot \mathbf{v}_D) \bar{\phi} \right)} - \mathcal{D}(f)$}}_{\text{Dissipation}}
\end{aligned}
$}
\end{equation*}
\colorbox{jaxpurple!20}{\textbf{Kinetic Dynamics} (I, II, IV).} They describe the collisionless trajectories of particles in the background magnetic configuration. \textit{Parallel Advection} (I) represents motion along magnetic field lines, while \textit{Magnetic Drift} (II) accounts for curvature and $\nabla B$ drifts. The \textit{Mirror Term} (IV) models the force arising from the parallel gradient of the magnetic field magnitude, which leads to particle trapping.

\colorbox{jaxtealdark!20}{\textbf{Energy Drives} (V, VII, VIII).} The \textit{Equilibrium Drive} (V) originates from thermodynamic gradients ($\nabla F_M$), providing the energy source for turbulence. \textit{Field Drives} (VII and VIII) represent the linear coupling between the electrostatic potential and the background distribution.

\colorbox{jaxred!20}{\textbf{Nonlinear Advection} (III).} The \textit{Nonlinear Term} $E \times B$ represents advection of the gyro-averaged distribution function by the fluctuating electric field. $E \times B$ is responsible for the saturated turbulent state, and the energy transfer across different spatial scales which pushes back on mode growth.

\colorbox{jaxblue!15}{\textbf{Dissipation}.} A numerical dissipation term is added for stability: 4th-order upwinded dissipation in the parallel direction, centered 4th-order smoothing in velocity space, and spectral hyper-viscosity in the perpendicular dimensions.

\colorbox{gray!25}{\textbf{Neoclassical} (VI, omitted).} $-\mathbf{v}_D \cdot \nabla F_M$ is the coupling of the magnetic drift (curvature/$\nabla B$, Coriolis, centrifugal) to equilibrium gradients. It is active only for rotating or neoclassical plasmas and we omit it for simplicity.

\section{\texttt{gyaradax} Implementation}
\label{sec:gyaradax}
\texttt{gyaradax} solves the collisionless electrostatic gyrokinetic equations in the local flux-tube limit, supporting both adiabatic and kinetic electrons. We designed it to be a slim, modern and ML-friendly alternative to GKW, written in JAX with the core integrator and solver in about 3\,000 lines.

\subsection{Time Integration}
\texttt{gyaradax} is designed as a purely functional solver, where the simulation state is evolved through a chain of stateless transformations.
We use explicit fourth-order Runge-Kutta (RK4) as integrator.
Integration across sequential timesteps is fused with \texttt{jax.lax.scan}, reducing the Python interpreter overhead.
Spatial derivatives in the parallel and velocity coordinates are computed using 4th-order central and upwinded stencils. The perpendicular dimensions are resolved pseudospectrally with 3/2-rule dealiasing to prevent aliasing errors in the nonlinear $E \times B$ term.

\subsection{Performance}
All species-dependent coefficients (Bessel functions, Maxwellians, drift velocities, and fused finite-difference stencils) are precomputed once before the time loop, eliminating redundant computation across the 4 RHS evaluations required by each RK4 step.
Moreover, transitioning to JAX opens up performance optimizations that would be not always trivial to implement in Fortran.
Specifically, for \texttt{gyaradax} we employed two orthogonal performance axes: \emph{mixed precision} and \emph{custom CUDA kernels}.
Together, these optimizations yield a ${\sim}2{\times}$ speedup over the JAX backend at the same grid resolution (\cref{tab:performance_benchmarks}).

\paragraph{Mixed precision.}
We observe that 2D FFTs, spatial derivatives and IFFTs in the nonlinear Poisson bracket can be performed in Float32 without loss of physical fidelity. Linear terms, field solver, and the final forward FFT on the accumulated output remain in Float64. This halves the memory bandwidth of the 8 FFTs per RK4 step.

\paragraph{Z2Z packing.}
We reduce the inverse FFT count from 4 to 2 by exploiting the linearity of the DFT. Same-field spatial derivatives ($ik_x \hat{f}$, $ik_y \hat{f}$) are packed into a single complex-to-complex (C2C) transform, with a Hermitian symmetrization correction at $k_y{=}0$ to prevent channel leakage from gyro-averaging asymmetry (Appendix~\ref{sec:nonlinear_fft}).

\begin{figure}
    \centering
    \includegraphics[width=0.8\linewidth]{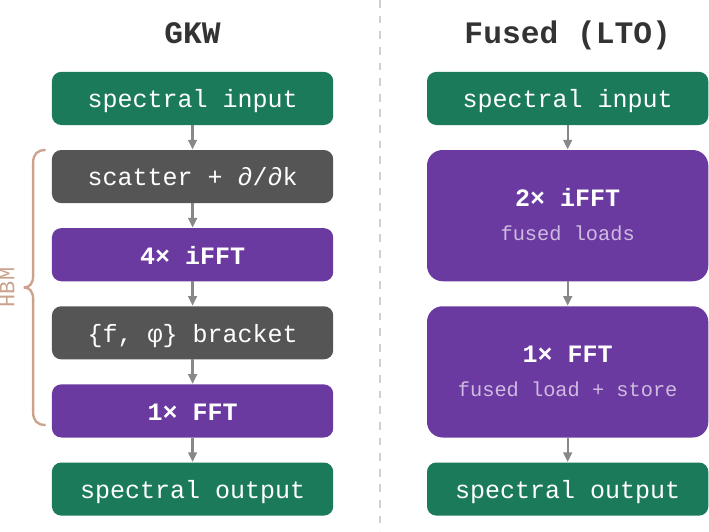}
    \caption{Nonlinear Poisson bracket. \textbf{Left:} GKW implementation with 4 inverse FFTs. \textbf{Right:} CUDA LTO callbacks fuse all pre- and post-FFT work, with Z2Z packing reducing the iFFTs to 2.}
    \label{fig:cufft_lto_fusion}
\end{figure}

\paragraph{CUDA backend.}
For the linear RHS, GKW and the JAX backend evaluate the nine-point parallel stencil, the five-point velocity stencil, and eight elementwise physics terms as separate operations fused by the compiler. Our custom CUDA kernel performs all of these in a single pass with strided addressing and velocity-axis tiling, eliminating the gather-and-predicate overhead introduced by XLA's general-purpose stencil lowering.
\cref{fig:cufft_lto_fusion} shows the optimized nonlinear Poisson bracket. Since XLA cannot fuse across cuFFT calls, we apply Link-Time Optimization (LTO) callbacks to embed the spectral derivative multiplication, gyro-averaging, bracket computation, and spectrum unpacking directly into cuFFT butterfly passes, avoiding intermediate HBM round-trips (Appendix~\ref{app:cuda_kernels}).

\subsection{Limitations and Differences to GKW}
The main shortcoming of \texttt{gyaradax} is that it can only handle electrostatic collisionless plasmas. It does not yet support electromagnetic perturbations, collisionality, Coriolis effects (Term~VI), or global gyrokinetics.
Time integration is limited to explicit RK4, which becomes slow for fast kinetic species due to tightening CFL constraint. GKW additionally provides the option of implicit or semi-implicit integrators for stiff parallel streaming.
As for geometry, we implement the analytic circular (Lapillonne) and s-alpha. Shaped plasmas require Miller parameterization or numerical MHD equilibria (EFIT/CHEASE), not yet in \texttt{gyaradax}.

We treat the 5D domain as dense arrays, instead of the sparse GKW format. Additionally, at the moment we do not support grid parallelism, unlike legacy codes that rely on global mutable state and complex MPI-based domain decomposition.
Adiabatic and kinetic electron models are available, supporting single-species (ion) and multi-species (ion + electron) simulations.
IO and simulation configurations are different formats but consistent with each other.
Our diagnostics suite matches GKW on the main fronts, with multi-species fluxes, spectra and growth rates.
%


\begin{figure*}[!t]
    \centering
    \begin{subfigure}{\linewidth}
        \centering
        \includegraphics[width=0.49\linewidth]{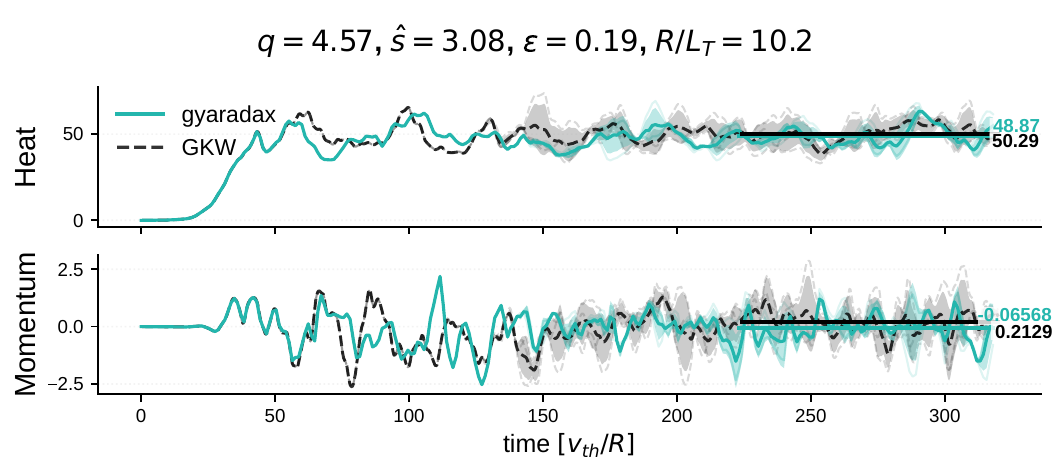}
        \hfill
        \includegraphics[width=0.49\linewidth]{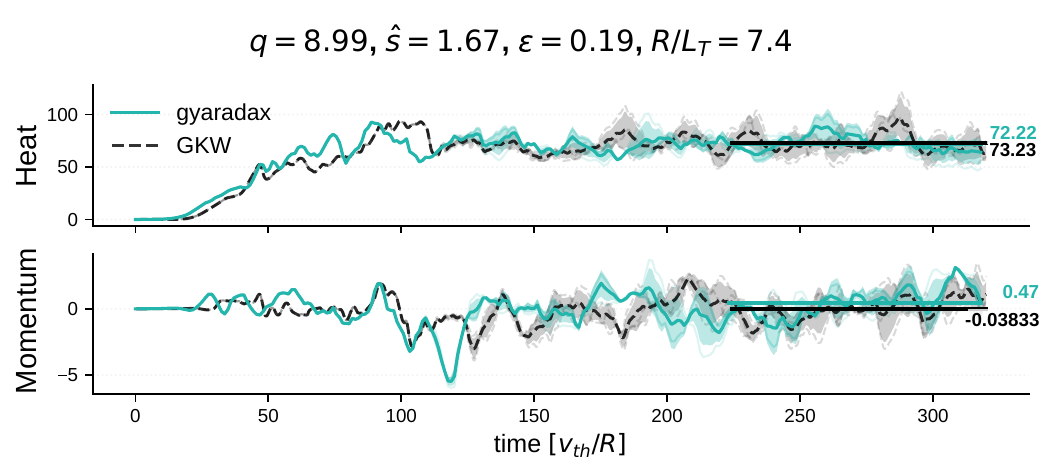}
        \label{fig:emp_fluxes}
    \end{subfigure}
    
    \vspace{-0.5em}
    
    \begin{subfigure}{\linewidth}
        \centering
        \includegraphics[width=0.5\linewidth]{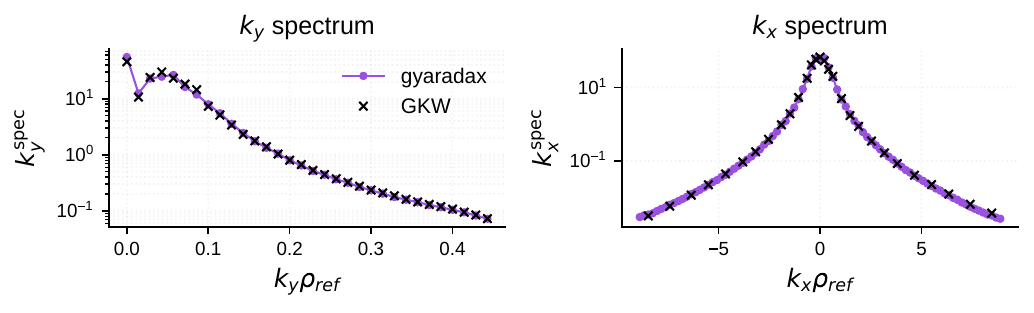}
        \hspace{-0.5em}
        \includegraphics[width=0.5\linewidth]{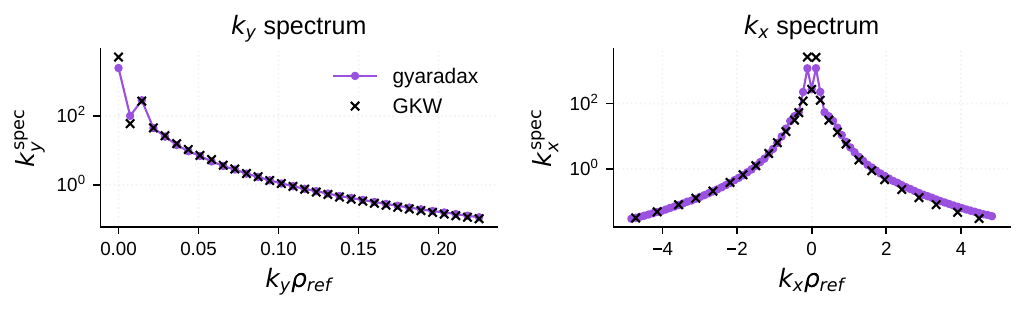}
        \label{fig:emp_spectra}
    \end{subfigure}
    \vspace{-1.5em}

    \caption{Empirical statistical validation of \texttt{gyaradax} against GKW for two adiabatic ITG equilibria.
    \textbf{Top:} heat and momentum flux time traces. Both codes coincide during linear growth, but once the nonlinear effects kick-in, trajectories diverge due to chaotic sensitivity. Three runs were performed for both \texttt{gyaradax} and GKW, with the variance across them reported as shaded bounds.
    \textbf{Bottom:} time-averaged $k_y$ and $k_x$ spectral profiles, with strong agreement on the slope and the peak location.
    }
    \label{fig:empirical}
\end{figure*}


\subsection{Vibecoding Setup}
\label{sec:vibecoding}
\texttt{gyaradax} was written and validated in a short time with substantial use of coding agents. We consider it relevant to add details on the workflow used.

\textbf{Workspace.}
In the preparation phase, we accomplish three things:
\begin{enumerate*}[label=\textbf{\roman*.}]
\item We provide utilities, such as GKW-specific I/O, parsing, as well as field solvers and flux integrals from \citet{galletti2026pinc}. The goal here is to help the agent target solver logic, instead of side tasks.
\item Some reference trajectories are included to understand the GKW configuration and output structure. Notably, these contain periodic $f$ dumps, simplifying debugging by providing the full state instead of integrated observables.
\item We wrote unit tests to symbolically and empirically verify alignment.
\end{enumerate*}
This \textit{empirical test-driven} cycle proved essential for the agent-led portion of the GKW rewrite.
We find that on hard tasks a measurable form of success is mandatory for vibecoding.

\textbf{Prompting.}
The first step in the prompt directed the agent to deliberate code ingestion and note taking, using the authors' knowledge of the solver to load core Fortran files and important manual pages fully into context.
After this phase, the task was to alternate between implementing solver logic, validating it with tests and incrementally adding new ones.
We explicitly decomposed the problem to first implement and validate the linear terms, then address the more involved nonlinear part (Term III).
We refined the initial instructions with meta-prompting, focusing on structure and imperative syntax better suited for modern LLMs \cite{zhang2025metapromptingaisystems}. The final version is provided in \cref{app:prompt}.

\textbf{Models.}
Initial attempts using Gemini 3.1 Pro \cite{gemini31pro_2026} via \texttt{gemini-cli} yielded unreasonable results over multiple parallel tries.
We had more success with GPT-5.3 CODEX \cite{openai_gpt53_codex_2026} for the core translation and Claude 4.6 Opus \cite{anthropic2026claude46opus} for the kinetic electron implementation.
%
For the final performance optimization we use a multi-agent workflow to generate and integrate custom CUDA kernels. This pipeline begins with a proposer/reviewer consensus loop. 
After a human review of the strategy, a lightweight model (Gemini 3 Flash, Claude Sonnet) iteratively implements the proposals. The verification pipeline handles compilation and benchmarking of the kernels against the JAX reference. See \cref{app:cuda_kernels} for a breakdown.
Finally, some opportunities for speedup in the early stages, such as precomputing the linear terms, were spotted and implemented entirely by Claude.


\section{Verification and Experiments}
\label{sec:verexp}
We validate \texttt{gyaradax} against GKW on nonlinear ITG turbulence, validate it on analytical benchmarks, and demonstrate AD through inverse problems and sensitivity analysis.


\subsection{Empirical Validation}
\label{sec:verification}
\begin{figure}[t]
    \centering
    \includegraphics[width=\columnwidth]{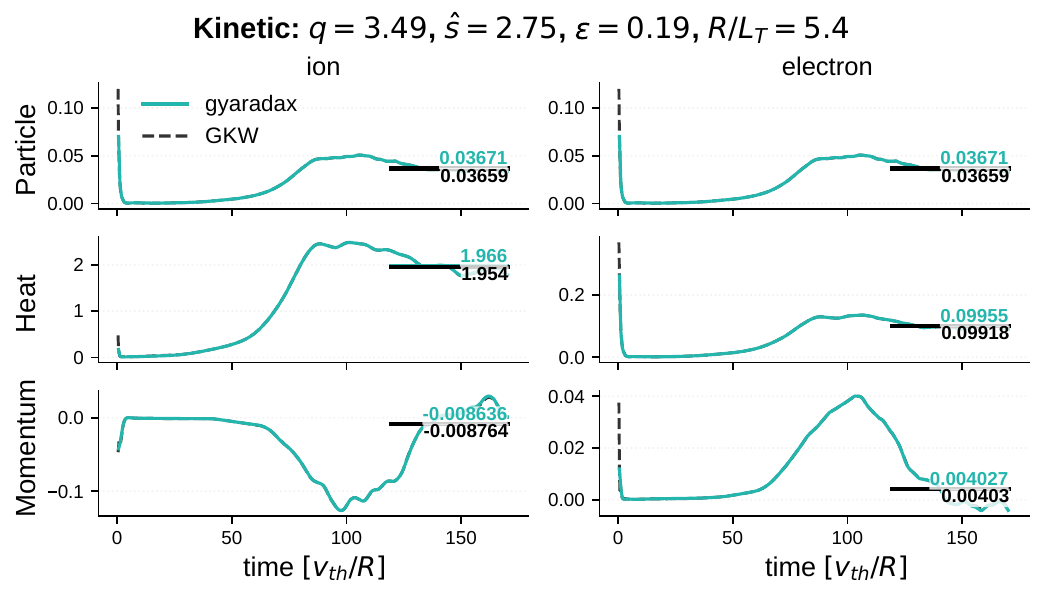}\\
    \includegraphics[width=\columnwidth]{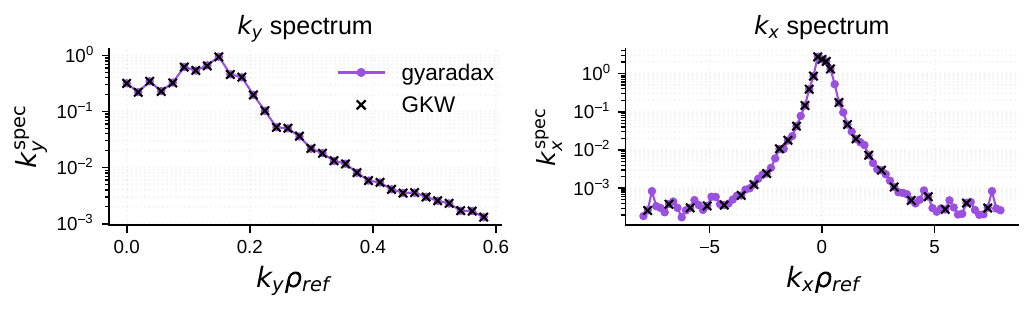}
    \caption{Kinetic electron validation.
    \textbf{Top:} ion (left) and electron (right) fluxes.
    \textbf{Bottom:} time-averaged $k_y$ and $k_x$ spectral profiles.}
    \label{fig:kinetic_fluxes}
\end{figure}

Pointwise verification of a chaotic system can be non representative: two runs of the same code from the same initial conditions will diverge once floating-point differences accumulate.
We therefore verify \texttt{gyaradax} through statistical agreement with GKW, covering both adiabatic \cite{paischer2025gyroswin} and kinetic electrons (\cref{fig:empirical,fig:kinetic_fluxes}).
All test cases use a standard grid ($N_{v_{\parallel}} = 32$, $N_\mu = 8$, $N_s = 16$, $N_{k_x} = 85$, $N_{k_y} = 32$).
The adiabatic cases use $\textrm{d}t = 0.01$, while kinetic electron cases use adaptive timestepping with $\textrm{d}t \approx 2.1 \times 10^{-3}$ (CFL).
We evaluate transport fluxes, wavenumber spectra, and per-$k_y$ growth rates.
To illustrate the chaotic sensitivity, we perform 3 runs per configuration from identical initial conditions, for both \texttt{gyaradax} and GKW.
Additionally, to quantify the statistical properties, agreement across a validation set of 46 unstable adiabatic runs is performed. Individual traces are in the Appendix \cref{fig:appx_fluxes,fig:appx_spectra}.

\textbf{Flux Traces.}
Transport fluxes are the primary observables for evaluating turbulence saturation and macroscopic plasma confinement.
For adiabatic simulations, heat flux is the most meaningful one, while particle flux vanishes identically.
In \cref{fig:empirical} (top), we compare the time evolution of ion heat and momentum fluxes. We plot an ensemble of three identical runs for both \texttt{gyaradax} and GKW to account for the chaotic sensitivity of the system.
In the linear growth phase both traces match, but after the onset of turbulence trajectories diverge. This is due to the accumulation of microscopic floating-point variations. For \texttt{gyaradax} this effect is amplified by the non-deterministic GPU optimization.
Despite these expected locally divergent phase spaces, the statistical behavior is similar, with time-averaged flux levels converging to the GKW baseline, confirming that the long-term turbulent transport is correctly reproduced.
Across conditions, the grand mean heat flux is close ($\bar{Q}_\texttt{gyaradax} = 90.9$, $\bar{Q}_\text{GKW} = 91.3$), with a rMAE of $0.14$.

\textbf{Spectra.}
The spectral energy distribution across perpendicular wavenumbers characterizes the turbulent cascade.
We compare the spectra $k_y^{\text{spec}} = \sum_{s, k_x} |\hat{\phi}(s, k_x, k_y)|^2$ and $k_x^{\text{spec}} = \sum_{s, k_y} |\hat{\phi}(s, k_x, k_y)|^2$, which measure how energy is distributed across binormal and radial scales respectively.
\cref{fig:empirical} (bottom) shows alignment of both time-averaged profiles, despite local differences in microstates.
\Cref{tab:spectra} summarizes the spectral agreement across all configurations.

\begin{table}[H]
\centering
\small
\caption{Spectral validation across 46 ITG equilibria (mean $\pm$ std).}
\label{tab:spectra}
\resizebox{\columnwidth}{!}{
\begin{tabular}{l c c c}
\toprule
 & KS statistic & Pearson $r$ & Log rel.\ $L_2$ \\
\midrule
$k_y^{\text{spec}}$ & $0.073 \pm 0.049$ & $0.990 \pm 0.018$ & $0.073 \pm 0.042$ \\
$k_x^{\text{spec}}$ & $0.055 \pm 0.035$ & $0.952 \pm 0.068$ & $0.076 \pm 0.042$ \\
\bottomrule
\end{tabular}
}
\end{table}

\textbf{Growth Rates.}
Growth rates $\gamma(k_y)$ measure the linear instability drive. In the nonlinear regime they converge to zero, showing the balance between linear drive and nonlinear saturation.
Across the validation set, both codes converge to $\bar\gamma \approx 0$ ($\bar\gamma_\texttt{gyaradax} \approx 9.2 \times 10^{-4}$ and $\bar\gamma_\text{GKW} \approx -1.8 \times 10^{-4}$), as expected for a balanced turbulent state.

\textbf{Kinetic Electrons.}
We additionally verify the solver on three kinetic electron configurations, with two species (ions and electrons).
Due to faster perturbations, adaptive CFL timestepping is required (constrained by the electron thermal velocity $v_{th,e}/v_{th,i} \approx 60$).
Per-species fluxes and $k_y$/$k_x$ spectra match the GKW reference (\cref{fig:kinetic_fluxes}). 

\subsection{Analytical Benchmarks}
\label{sec:analytical}
\texttt{gyaradax} is verified on two standard benchmarks with analytical solutions. Details in \cref{app:analytical_details}.

\begin{figure}[H]
    \centering
    \includegraphics[width=\columnwidth]{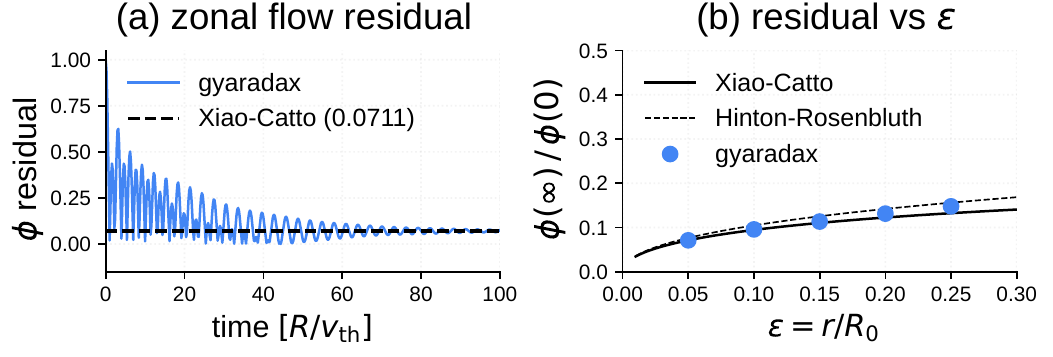}
    \vspace{-0.8em}
    \caption{Rosenbluth-Hinton zonal flow test at $q{=}1.3$. (a)~Zonal potential residual over time at $\varepsilon{=}0.05$. (b)~Residual at varying $\varepsilon$.}
    \label{fig:rh}
\end{figure}

\textbf{Rosenbluth-Hinton Test.}
The RH test is a sensitive end-to-end test of the field solver and the linear dynamics.
The zonal flow ($k_\zeta{=}0$) excites a geodesic acoustic mode that oscillates and damps via collisionless Landau damping.
The residual zonal potential is given analytically by $\phi(\infty)/\phi(0) = 1/(1 + q^2 \Theta/\varepsilon^2)$, where $\Theta(\varepsilon)$ captures finite-$\varepsilon$ corrections \cite{Rosenbluth,xiao2006plasma}.
At $q{=}1.3$, $\varepsilon{=}0.05$, $\hat{s}{=}0.16$, 
\texttt{gyaradax} converges to a residual of $0.0711$, matching the prediction (\cref{fig:rh}a).
A scan over $\varepsilon$ at fixed $q$ follows the analytical curve across the full range (\cref{fig:rh}b).

\begin{figure}[H]
    \centering
    \includegraphics[width=\columnwidth]{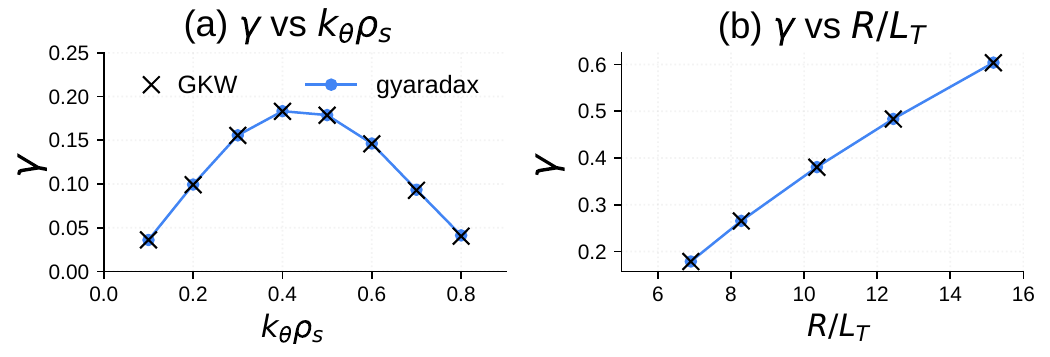}
    \vspace{-0.8em}
    \caption{Cyclone Base Case (linear). (a)~Growth rate vs $k_\theta\rho_s$ ($R/L_T{=}6.9$). (b)~Growth rate vs $R/L_T$ ($k_\theta\rho_s{=}0.5$).}
    \label{fig:cbc}
\end{figure}

\textbf{Cyclone Base Case.}
CBC \cite{cbc} is the standard linear benchmark. It tests the full linear operator in the regime where the ion temperature gradient (ITG) drives instability. The configuration used has adiabatic electrons, $q{=}1.4$, $\hat{s}{=}0.78$, $\varepsilon{=}0.19$, $R/L_n{=}2.2$ and s-alpha geometry.
\texttt{gyaradax} matches growth rates $\gamma(k_\theta\rho_s)$ at $R/L_T{=}6.9$ (\cref{fig:cbc}a) and the gradient dependence $\gamma(R/L_T)$ at $k_\theta\rho_s{=}0.5$ (\cref{fig:cbc}b).

\subsection{Experiments}
\label{sec:exp}

\textbf{Inverse Problem.}
\label{sec:inverse}
\begin{figure}[b]
    \includegraphics[height=0.5\linewidth]{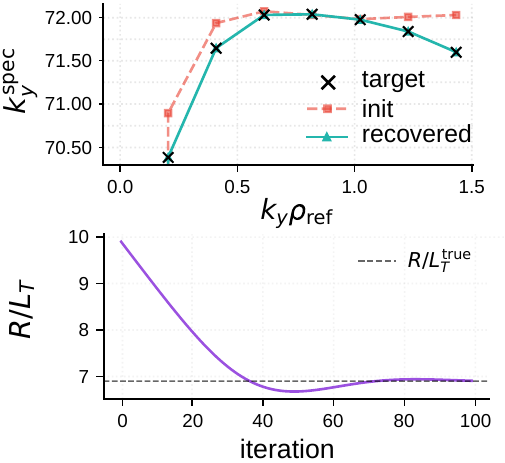}\hspace{-0.3em}
    \includegraphics[height=0.5\linewidth]{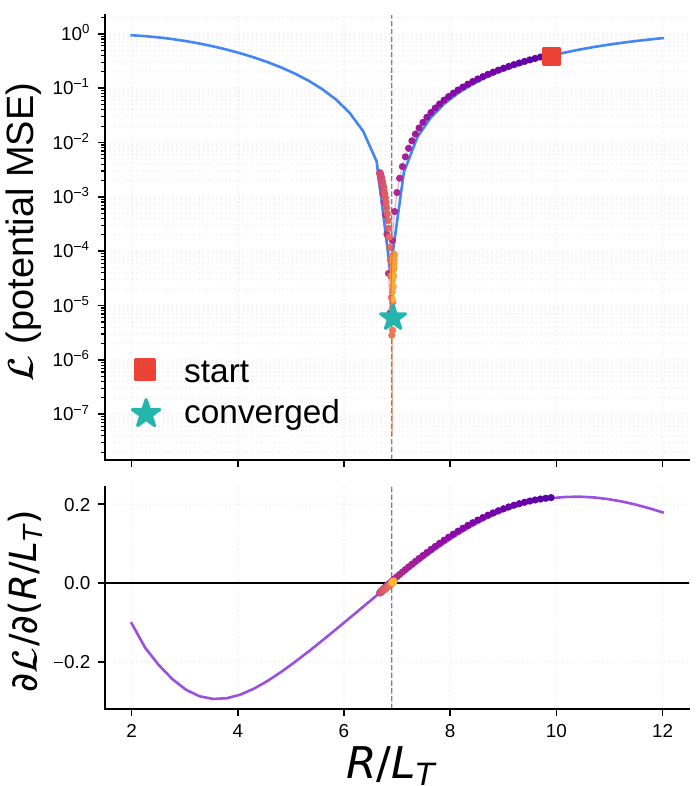}
    
    \caption{Recovery of $R/L_T$ from the electrostatic potential. 
    \textbf{Left:} initial, target ($\times$), and recovered $k_y$ spectra (top), $R/L_T$ convergence during iterations (bottom). 
    \textbf{Right:} loss landscape $\mathcal{L}(R/L_T)$ (top) and AD gradient (bottom), with the Adam trajectory overlaid.}
    \label{fig:inverse}
\end{figure}
A direct payoff of end-to-end differentiability is gradient-based inverse problems.
We demonstrate this by recovering the temperature gradient $R/L_T$ from a target electrostatic potential $\phi^*$.
We fix the equilibrium parameters ($q{=}1.4$, $\hat{s}{=}0.78$, $\varepsilon{=}0.19$, $R/L_n{=}2.22$) and use a reduced grid $(N_{v\parallel}, N_\mu, N_s, N_{k_x}, N_{k_y}) = (24, 8, 16, 9, 8)$.
The target $\phi^*$ is generated by running a linear simulation at $R/L_T^{\text{true}} = 6.9$ for 400 steps ($\mathrm{d}t=0.01$).
The loss is the potential energy mismatch
\begin{equation*}
    \mathcal{L}(R/L_T) = \frac{1}{N} \sum_{s, k_x, k_y} \bigl|\phi(s, k_x, k_y;\, R/L_T) - \phi^*\bigr|^2 \ ,
\end{equation*}
where $\phi$ is the result of the full forward solve. Reverse-mode AD provides exact $\partial\mathcal{L}/\partial(R/L_T)$ through the full 400-step integration.
Starting from $R/L_T^{(0)} = 10.0$, Adam \cite{kingma_adam_2015} converges to $R/L_T = 6.908$ (\cref{fig:inverse}).
The recovered potential matches the target $k_y^{\text{spec}}$ (\cref{fig:inverse}, top left), confirming that the optimizer recovers both the instability drive and the eigenmode structure.
The loss landscape (\cref{fig:inverse}, right), obtained by evaluating $\mathcal{L}$ and $\partial\mathcal{L}/\partial(R/L_T)$ over $R/L_T \in [2, 12]$, is smooth and unimodal, with a clean zero crossing of the gradient at the optimum.

\textbf{Sensitivity Analysis.}
\label{sec:sensitivity}
\begin{figure}[t]
    \centering
    \includegraphics[width=0.79\columnwidth]{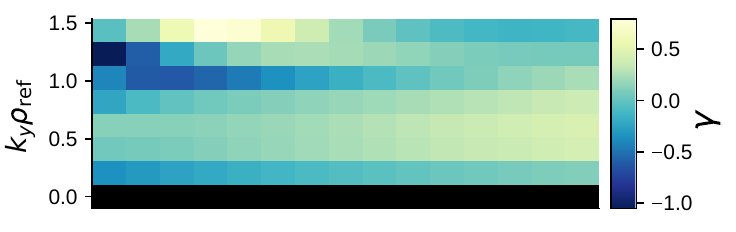}\\\vspace{-8px}
    \includegraphics[width=0.79\columnwidth]{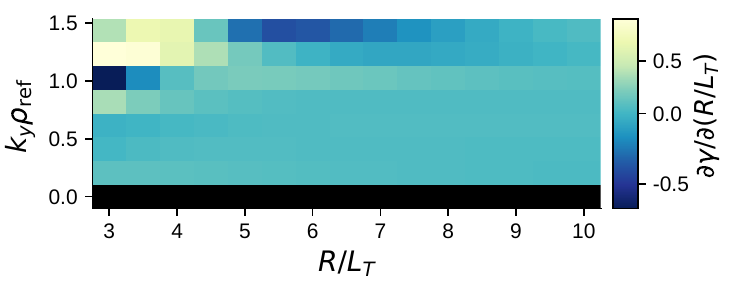}
    \caption{Growth rate sensitivity to $R/L_T$ ($k_y{=}0$ zonal mode is masked).
    \textbf{Top:} per-$k_y$ growth rate $\gamma(k_y, R/L_T)$.
    \textbf{Bottom:} sensitivity $\partial\gamma / \partial(R/L_T)$.}
    \label{fig:sensitivity}
\end{figure}
Sensitivity analysis measures responses to parameter changes, informing design and uncertainty quantification.
Finite differences require $2N$ evaluations for $N$ parameters and suffer from step-size errors.
In contrast, reverse-mode AD computes exact gradients of scalar losses in one backward pass. For vector-valued outputs like $\gamma(k_y)$, \texttt{jax.jacrev} computes the full Jacobian in $N_{k_y}$ passes (one per output dimension).
Using the inverse problem setup, we scan $R/L_T \in [3, 10]$ across 15 values, computing $\gamma(k_y)$ and its Jacobian $\partial\gamma / \partial(R/L_T)$.
All modes are damped ($\gamma < 0$) on this grid (\cref{fig:sensitivity}, top).
Increasing $R/L_T$ reduces the damping rate, with high-$k_y$ modes at low drive being the most responsive (\cref{fig:sensitivity}, bottom).

\subsection{Performance Comparison vs GKW}
The performance of \textit{gyaradax} with different settings is shown in \cref{tab:performance_benchmarks}, showing significant performance gains over GKW in both adiabatic and kinetic electron cases. The CUDA backend requires approximately 1.2 GB additional GPU memory beyond the JAX implementation, as the fused cuFFT kernel pre-allocates its own workspace via \textit{cudaMalloc}.
GKW is run on an AMD EPYC 9754 128-Core Processor with 1TB of RAM, with 64 and 128 processes for adiabatic and kinetic respectively. All \texttt{gyaradax} benchmarks are performed on a single Nvidia Blackwell B300 SXM6 GPU (275GB variant).
Overall, JAX with Z2Z packing and mixed precision leads to a $5.4\times$ speedup compared to GKW, and when using CUDA kernels it jumps to $10.5\times$.

\begin{table}[ht]
\centering
\caption{Benchmarks comparing GKW against \texttt{gyaradax} variants. All runs use the same grid resolution $(32, 8, 16, 85, 32)$. Note that GKW uses main memory (RAM) and \texttt{gyaradax} uses GPU memory (VRAM). Configurations denote the computational backend and numerical precision of the nonlinear term.}
\label{tab:performance_benchmarks}
\vspace{0.15cm}
\resizebox{\columnwidth}{!}{%
\begin{tabular}{@{} l c c c @{}}
\toprule
\textbf{Solver Configuration} & \textbf{Steps/s} & \textbf{Speedup} & \textbf{Mem (GB)} \\
\midrule
\multicolumn{4}{c}{Adiabatic Electrons} \\
\midrule
GKW (DP)                   & 5.75  & -- & 18.0 \\
\texttt{gyaradax} (JAX, DP)  & 12.49 & 2.17$\times$  & 9.4  \\
\texttt{gyaradax} (JAX, MP) & 30.89 & 5.37$\times$  & 9.4  \\
\texttt{gyaradax} (CUDA, DP)& 16.6  & 2.89$\times$  & 10.6 \\
\texttt{gyaradax} (CUDA, MP)& 60.54 & \textbf{10.53}$\bm{\times}$ & 10.6 \\
\midrule
\multicolumn{4}{c}{Kinetic Electrons} \\
\midrule
GKW (DP)                   & 3.43  & -- & 38.8 \\
\texttt{gyaradax} (JAX, DP)  & 6.21  & 1.81$\times$  & 17.6 \\
\texttt{gyaradax} (JAX, MP) & 15.19 & 4.43$\times$  & 17.6 \\
\texttt{gyaradax} (CUDA, DP) & 8.13   & 2.36$\times$  & 18.9 \\
\texttt{gyaradax} (CUDA, MP)& 28.16 & \textbf{8.21}$\bm{\times}$  & 18.9 \\
\bottomrule
\end{tabular}%
}
\end{table}


\section{Conclusion and Future Work}
\texttt{gyaradax} is a minimal and fully differentiable JAX implementation of local flux-tube gyrokinetics, supporting both adiabatic and kinetic electrons.
Development was heavily supported by modern agentic workflows, with human oversight.
Our implementation is formally correct and achieves numerical parity with GKW, while offering native hardware acceleration and easy integration with ML pipelines.
The availability of an end-to-end differentiable gyrokinetic solver can open new avenues for plasma research, from automated parameter optimization \cite{McGreivy_2021,joglekar2026differentiableprogrammingplasmaphysics} to the development of physics-informed surrogate models \cite{um2021solverinthelooplearningdifferentiablephysics,Karniadakis2021}.
Moreover, the lightweight and accessible nature of \texttt{gyaradax} lowers the entry barrier of an otherwise highly specialized code, making extensions, distribution and maintenance substantially easier.

\textbf{Future Work.}
Currently, \texttt{gyaradax} is limited to electrostatic collisionless plasmas. Future work will incorporate electromagnetic perturbations ($A_\parallel$, $B_\parallel$) and collisionality. As problem sizes grow significantly when faster species are introduced, support for grid parallelism via GPU sharding \cite{xu2021gspmdgeneralscalableparallelization} is required.
Another promising direction is fully spectral solvers \cite{CANDY201673}, where all spatial dimensions are resolved in Fourier/spectral space without finite-difference stencils. Such formulations are usually better suited for GPU acceleration, as they replace stencils with global spectral transforms that map naturally onto batched matrix operations.
On the verification side, future work includes extending the analytical test suite to kinetic-electron benchmarks and electromagnetic cases from \citet{PEETERS_GKW_2009}.
Ultimately, we present our work as a proof-of-concept for the rapid agent-assisted development of modern high-performance scientific software.

\section*{Acknowledgments}
We thank \textbf{William Hornsby} for discussions on GKW and formal verification.

The ELLIS Unit Linz, the LIT AI Lab, the Institute for Machine Learning, are supported by the Federal State Upper Austria. We thank the projects FWF AIRI FG 9-N (10.55776/FG9), AI4GreenHeatingGrids (FFG- 899943), Stars4Waters (HORIZON-CL6-2021-CLIMATE-01-01), FWF Bilateral Artificial Intelligence (10.55776/COE12). We thank NXAI GmbH, Silicon Austria Labs (SAL), Merck Healthcare KGaA, GLS (Univ. Waterloo), T\"{U}V Holding GmbH, Software Competence Center Hagenberg GmbH, dSPACE GmbH, TRUMPF SE + Co. KG.

\bibliographystyle{icml2026}
\bibliography{references}

\newpage
\appendix
\crefalias{section}{appendix}
\onecolumn
\section{Parameters}

\subsection{Phase-Space Variables}
\renewcommand{\arraystretch}{1.4}
\begin{longtable}{p{0.125\textwidth} p{0.2\textwidth} p{0.525\textwidth}}
\toprule
\textbf{Symbol} & \textbf{Code Key} & \textbf{Description} \\
\midrule
\endfirsthead
\toprule
\textbf{Symbol} & \textbf{Code Key} & \textbf{Description} \\
\midrule
\endhead
\bottomrule
\endfoot
$\delta f_a$ & \texttt{df} & Perturbed distribution function for species $a$. Shape: $(N_{v\parallel}, N_\mu, N_s, N_{kx}, N_{ky})$ for adiabatic; $(N_{sp}, N_{v\parallel}, N_\mu, N_s, N_{kx}, N_{ky})$ for kinetic electrons. \\
$\phi, \langle\phi\rangle$ & \texttt{phi, gyro\_phi} & Electrostatic and gyro-averaged potential ($J_0 \phi$). Shape: $(N_s, N_{kx}, N_{ky})$. \\
$s$ & \texttt{sgrid} & Field-line coordinate, $s \in [-0.5, 0.5]$ for \texttt{nperiod}$=1$ (one poloidal transit). \\
$v_\parallel, \mu$ & \texttt{vpgr, mugr} & Parallel velocity (uniform grid) and magnetic moment ($\mu = v_\perp^2/2$, uniform in $v_\perp$). \\
$k_x, k_y$ & \texttt{kxrh, krho} & Radial and binormal spectral wavevectors, normalized to $\rho_{ref}$. \\
$\Delta s, \Delta v_\parallel$ & \texttt{sgr\_dist, dvp} & Parallel and velocity grid spacing. \\
$w_s, w_v, w_\mu$ & \texttt{ints, intvp, intmu} & Phase-space integration weights for $s$, $v_\parallel$, and $\mu$. The $\mu$ weights are $2\pi v_\perp \Delta v_\perp$ (cylindrical Jacobian). \\
$P_{k_y}$ & \texttt{parseval} & Parseval normalization: 1 for $k_y=0$, 2 for $k_y > 0$ (half-spectrum convention accounts for the conjugate mode). \\
$\Delta t$ & \texttt{dt} & Small timestep for RK4 integration. May vary per step when CFL-adaptive. \\
\bottomrule
\end{longtable}

\subsection{Species Parameters}

\begin{longtable}{p{0.125\textwidth} p{0.2\textwidth} p{0.525\textwidth}}
\toprule
\textbf{Symbol} & \textbf{Code Key} & \textbf{Description} \\
\midrule
\endfirsthead
\midrule
\endhead
\bottomrule
\endfoot
$m_a, T_a$ & \texttt{mas, tmp} & Species mass and temperature, normalized to reference. Scalar (adiabatic) or array of length $N_{sp}$ (kinetic). \\
$n_a, Z_a$ & \texttt{de, signz} & Species density and charge number. Same shape convention as mass/temperature. \\
$R/L_T, R/L_n$ & \texttt{rlt, rln} & Inverse temperature and density gradient scale lengths. Per-species for kinetic electrons. \\
$v_{th,a}/v_{th,ref}$ & \texttt{vthrat} & Thermal velocity ratio $\sqrt{T_a/m_a}$ relative to reference species. \\
$F_{M,a}, J_0$ & \texttt{fmaxwl, bessel} & Background Maxwellian and Bessel $J_0(k_\perp \rho_a)$. Per-species in kinetic mode. \\
$\Gamma_0$ & \texttt{gamma} & FLR kernel $I_0(b_a) e^{-b_a}$ with $b_a = \tfrac{1}{2}(m_a v_{th,a} k_\perp / Z_a B)^2$. Computed via \texttt{jax.scipy.special.i0e} for numerical stability. \\
\bottomrule
\end{longtable}

\subsection{Geometry and Magnetic Equilibrium}
\label{app:geometry}

The gyrokinetic equation is formulated in a field-aligned coordinate system $(\psi, \zeta, s)$, where $\psi$ labels the flux surface (radial direction), $\zeta$ is the field-line label (binormal direction), and $s$ follows the magnetic field line (parallel direction). The \textit{geometry} encodes how this abstract coordinate system maps to physical space: it provides the covariant metric tensor $g_{ij}$ (for $k_\perp^2$ and perpendicular gradients), the magnetic field strength $B(s)$ (for mirror forces, gyro-averaging, and drifts), and a set of derived tensors that enter the gyrokinetic equation as advection coefficients.

\texttt{gyaradax} supports a standalone circular equilibrium model (Lapillonne, translated from GKW \texttt{geom.f90}) that computes all geometry arrays from three equilibrium scalars $(q, \hat{s}, \varepsilon)$ plus grid resolution parameters, eliminating the need for precomputed geometry files. The continuous geometry functions (B-field, metric, drift tensors) are implemented in JAX and are differentiable with respect to the equilibrium parameters; the discrete topology (mode labels, parallel boundary connectivity) uses numpy and is detached from the gradient graph via \texttt{jax.lax.stop\_gradient}.

\textbf{Equilibrium parameters.}
The safety factor $q$ controls how tightly wound the field lines are (large $q$ = nearly toroidal lines, weak poloidal field). The magnetic shear $\hat{s} = (r/q)\,dq/dr$ measures how $q$ varies radially and drives the spectral mode connectivity: adjacent $k_x$ modes couple across the parallel boundary with shift $\Delta k_x = 2\pi\hat{s}\,k_y$. The inverse aspect ratio $\varepsilon = r/R_0$ sets the strength of toroidal effects: the outboard midplane ($\theta=0$) sees a weaker field than the inboard side ($\theta=\pi$), producing trapped-particle physics and the ballooning structure characteristic of tokamak turbulence.

In the circular model, the magnetic field strength is:
\begin{equation}
B(s) = \frac{\delta}{1 + \varepsilon\cos\theta(s)}, \qquad \delta = \sqrt{1 + \frac{\varepsilon^2}{q^2(1-\varepsilon^2)}}
\label{eq:bn_circ}
\end{equation}
where $\theta(s)$ is obtained by inverting $\theta + \varepsilon\sin\theta = 2\pi s$.

\textbf{Metric tensor.}
The metric components $g_{\psi\psi}$, $g_{\psi\zeta}$, $g_{\zeta\zeta}$, $g_{\psi s}$, $g_{\zeta s}$, $g_{ss}$ define how coordinate distances map to physical distances. The solver uses these in the perpendicular wavenumber calculation:
\begin{equation}
k_\perp^2 = k_y^2 g_{\zeta\zeta} + 2 k_x k_y\, g_{\psi\zeta} + k_x^2 g_{\psi\psi}
\end{equation}
The cross-term $g_{\psi\zeta}$ (the ``dzetadeps'' coupling) encodes the integrated magnetic shear: it vanishes at the outboard midplane and grows along the field line, tilting eddies via $k_{x,\text{loc}} = k_x + k_y\, g_{\psi\zeta}$.
The $d\zeta/d\varepsilon$ computation uses a branch-tracked $\arctan$ with a finite-$\varepsilon$ correction (GKW \texttt{geom.f90} lines 1492--1511). This correction introduces an inherent $\sim 0.1\%$ model-level approximation that propagates into all $\zeta$-direction drift tensors ($D_\zeta$, $H_\zeta$, $I_\zeta$), while the radial ($\psi$) components remain accurate to $< 10^{-4}$.

\textbf{Drift tensors.}
All derivatives ($dB/d\psi$, $dR/d\psi$, $dZ/d\psi$, etc.) are first computed in $(\psi, \theta)$ space, then transformed to $(\psi, s)$ coordinates via the Jacobian:
\begin{equation}
f^{(s)}_\psi = f^{(\theta)}_\psi - \frac{\sin\theta}{1+\varepsilon\cos\theta}\, f_\theta, \qquad
f_s = \frac{2\pi}{1+\varepsilon\cos\theta}\, f_\theta
\end{equation}
The $E$-tensor (ExB operator) is built from the antisymmetric cofactors of the first two metric rows, scaled by $\pi\, dp_f/d\psi / B^2$ where $dp_f/d\psi = \varepsilon/(q\sqrt{1-\varepsilon^2})$. The curvature drift tensor is then:
\begin{equation}
D_j = \frac{-2\bigl(E_{j,\psi}\,\partial_\psi B + E_{j,s}\,\partial_s B\bigr)}{B}
\end{equation}
$D_\psi$ drives radial transport (proportional to $\sin\theta$, strongest at top/bottom); $D_\zeta$ drives binormal transport (proportional to $\cos\theta$ + shear corrections, strongest at the outboard midplane). They enter Term~II of the gyrokinetic equation as $-i(k_x D_\psi + k_y D_\zeta)(v_\parallel^2 + \mu B)\,\delta f$.

The $H$-tensor (Coriolis drift) uses $dZ/d\psi$ and $dZ/ds$ with metric coupling; the $I$-tensor (centrifugal drift) uses $E \cdot \nabla R$ scaled by $2R$. Both are stored for future rotation physics.

\begin{longtable}{p{0.1\textwidth} p{0.1\textwidth} p{0.7\textwidth}}
\toprule
\textbf{Symbol} & \textbf{Code Key} & \textbf{Description} \\
\midrule
\endfirsthead
\midrule
\endhead
\bottomrule
\endfoot
$q$ & \texttt{q} & Safety factor. Controls field-line winding and parallel domain extent. \\
$\hat{s}$ & \texttt{shat} & Magnetic shear $(r/q)\,dq/dr$. Drives mode connectivity across the parallel boundary. \\
$\varepsilon$ & \texttt{eps} & Inverse aspect ratio $r/R_0$. Controls trapped-particle fraction and ballooning. \\
$B(s)$ & \texttt{bn} & Magnetic field magnitude along the field line, normalized to $B_{ref}$. \\
$\mathcal{F}$ & \texttt{ffun} & Parallel streaming coefficient: $\mathcal{F} = b_{ups}/B$ where $b_{ups} = 1/(2\pi q\sqrt{1-\varepsilon^2})$. \\
$\mathcal{G}$ & \texttt{gfun} & Mirror force: $\mathcal{G} = \mathcal{F} \cdot (\partial_s B)/B$. Drives particle trapping. \\
$g_{ij}$ & \texttt{little\_g} & Perpendicular metric: stores $(g_{\zeta\zeta}, g_{\psi\zeta}, g_{\psi\psi})$ for $k_\perp^2$. \\
$\mathcal{D}_j$ & \texttt{dfun} & Curvature + $\nabla B$ drift (3 components: $\psi$, $\zeta$, $s$). \\
$\mathcal{E}_{\psi\zeta}$ & \texttt{efun} & ExB geometric factor: $-E_{0,1}(s)$, varies along the field line through the metric and $B^2$. \\
$\mathcal{H}_j$ & \texttt{hfun} & Coriolis drift tensor (3 components). For rotation physics. \\
$\mathcal{I}_j$ & \texttt{ifun} & Centrifugal drift tensor (3 components). For rotation physics. \\
$B_t/B$ & \texttt{bt\_frac} & Toroidal fraction of the total field $= 1/\delta$. \\
$R$ & \texttt{rfun} & Major radius along the field line: $R = 1 + \varepsilon\cos\theta$. \\
$k_{th}$ & \texttt{kthnorm} & Binormal wavenumber normalization: $k_{th} = \sqrt{g_{\zeta\zeta}(\theta=0)}$. \\
\bottomrule
\end{longtable}

\section{Functions} \label{sec:functions}
\renewcommand{\arraystretch}{1.4}
\begin{longtable}{p{0.3\textwidth} p{0.65\textwidth}}
\toprule
\textbf{Function} & \textbf{Description} \\
\midrule
\endfirsthead
\toprule
\textbf{Function} & \textbf{Role} \\
\midrule
\endhead
\bottomrule
\endfoot
\texttt{compute\_geometry} & Computes all geometry arrays from $(q, \hat{s}, \varepsilon)$. Continuous quantities (B-field, metric, drifts) use JAX and are differentiable; discrete topology (mode labels, connectivity) uses numpy. \\
\texttt{linear\_precompute} & Precomputes all species-dependent coefficients: Bessel functions $J_0$, Maxwellians $F_M$, drift velocities, advection speeds, and fused stencils. For kinetic electrons, all arrays gain a leading species dimension $(N_{sp}, \ldots)$. Called once before the time loop. \\
\texttt{\_linear\_rhs\_core} & Inner linear RHS for a single species (5D). Evaluates Terms I (streaming), II (drift), IV (mirror), V (drive), VII (Landau), VIII (drift drive), plus all dissipation operators. Uses precomputed fused stencils for parallel derivatives. \\
\texttt{nonlinear\_term\_iii} & Pseudospectral ExB advection via 2D FFTs with 3/2-rule dealiasing. Vectorized over the parallel grid via \texttt{jax.vmap}. \\
\texttt{calculate\_phi} & Unified phi solver: dispatches to adiabatic or kinetic path based on \texttt{params.adiabatic\_electrons}. Adiabatic path includes zonal flux-surface-averaged correction. \\
\texttt{estimate\_timestep} & Combined CFL: $\min(\Delta t_\text{NL}, \Delta t_\text{lin})$. Nonlinear CFL from real-space $|\nabla\phi|$; linear CFL from von Neumann analysis of streaming, trapping, and dissipation stencils. For kinetic electrons, the field CFL (electrostatic Alfv\'en frequency) is additionally included. \\
\texttt{gkstep\_single} & Single RK4 step. Dispatches to adiabatic or kinetic paths via static branching on \texttt{params.adiabatic\_electrons}. The kinetic path uses \texttt{jax.vmap} over species for both linear and nonlinear terms. \\
\texttt{gksolve} & Multi-step driver using \texttt{jax.lax.scan}. Supports fixed and CFL-adaptive timestep modes. Returns final $(\delta f, \phi, \text{fluxes}, \text{state})$. \\
\texttt{gk\_init} / \texttt{gksimulate} & High-level entry points: \texttt{gk\_init} creates initial conditions and state; \texttt{gksimulate} runs the full simulation with checkpointing and diagnostics. \\
\texttt{calculate\_fluxes\_kinetic} & Per-species transport fluxes $(Q_a^{particle}, Q_a^{heat}, Q_a^{momentum})$. Returns $(N_{sp}, 3)$ array. \\
\bottomrule
\end{longtable}
\renewcommand{\arraystretch}{1.0}

\section{Implementation Details}

\subsection{Spatial and Temporal Discretization}
Parallel ($s$) derivatives use 4th-order finite differences with 9-point stencils. The stencil class at each grid point is determined by \texttt{pos\_par\_grid\_class} $\in \{-2,-1,0,1,2\}$, encoding proximity to the open boundary. Interior points use centered 4th-order; boundary-adjacent points use upwinded 2nd-order. Upwind direction is selected by the sign of $v_\parallel$ (for streaming) or $\partial(J_0\phi)/\partial s$ (for Term~VII). Stencil coefficients and upwind selection are precomputed and fused into \texttt{s\_total\_upar} and \texttt{s\_total\_t7} arrays, eliminating per-step branching inside the RK4 loop.

Parallel velocity ($v_\parallel$) derivatives use 4th-order centered stencils with zero-padding at the grid boundaries (no-flux condition).

Perpendicular coordinates are resolved in spectral space. The nonlinear ExB term uses 2D real-to-complex FFTs with 3/2-rule zero-padding for dealiasing. The dealiased grid size is chosen to have small prime factors ($\leq 7$), with a preference for powers of two. Physical modes are mapped to the FFT storage layout via the \texttt{jind} index array.

Time integration uses the classical 4th-order Runge-Kutta scheme. Each step requires four evaluations of the full RHS (phi solve + linear terms + nonlinear term). An optional CFL-adaptive mode estimates the timestep as the minimum of: (i) the nonlinear CFL from the maximum real-space ExB velocity gradient, (ii) a von Neumann stability analysis of the streaming and trapping stencils, and (iii) for kinetic electrons, a field CFL based on the electrostatic Alfv\'{e}n frequency $\omega \sim k_\perp \sqrt{m_e / m_i}\, v_{th,i} / (q\, \Delta s\, B)$. The timestep is updated with one-step lag; the initial step uses a conservative linear-only estimate.

\subsection{Mode Connectivity and Open Boundaries}
Magnetic shear couples adjacent $k_x$ modes across the parallel boundary: a mode at $(k_x, k_y)$ that exits the field-line domain at $s = +\tfrac{1}{2}$ continues at $(k_x + \Delta k_x,\, k_y)$ with $\Delta k_x = 2\pi\hat{s}\,k_y$. This is discretized by grouping $k_x$ modes into chains spaced $\texttt{ikxspace} = \lfloor 2\pi\hat{s}\,\Delta k_y / \Delta k_x \rceil$ apart.
The \texttt{mode\_label} array assigns the same integer label to all $k_x$ indices in a chain for each $k_y$; \texttt{ixplus}/\texttt{ixminus} store the connected $k_x$ index in the positive/negative $s$-direction ($-1$ = open boundary, no connection).

For the zonal mode ($k_y = 0$), the parallel domain is periodic (each $k_x$ maps to itself). For $k_y > 0$, modes at the end of a chain see an open boundary: the stencil class \texttt{pos\_par\_grid\_class} $\in \{-2, -1, 0, 1, 2\}$ is set to $\pm 2$ and $\pm 1$ at the first/second grid points adjacent to an open boundary, switching the parallel stencil from 4th-order centered to lower-order upwinded. Interior points ($\texttt{class} = 0$) use centered stencils.

The full connectivity is precomputed into \texttt{s\_shift}, \texttt{kx\_shift}, and \texttt{valid\_shift} arrays of shape $(2L{+}1, N_s, N_{k_x}, N_{k_y})$ with $L = 4$ (the stencil half-width). These encode, for every stencil offset $\delta s \in [-4, 4]$, the target $(s, k_x)$ index and whether the shifted point is in-grid. The parallel derivative is then a single gather-and-dot operation without any per-step branching on boundary conditions.

\subsection{Multi-Species Architecture}
The kinetic electron extension adds a leading species dimension to the distribution function: $\delta f$ becomes $(N_{sp}, N_{v\parallel}, N_\mu, N_s, N_{kx}, N_{ky})$. All species-dependent precomputed arrays (Bessel functions, Maxwellians, drift velocities, fused stencils) similarly gain the species axis.

The RHS computation is vectorized over species using \texttt{jax.vmap}. Each species receives its own precomputed coefficients while sharing the same electrostatic potential $\phi$ from multi-species quasi-neutrality. The fused stencils have shape $(9, N_{sp}, N_{v\parallel}, \ldots)$; under \texttt{vmap} over species axis 1, each call sees $(9, N_{v\parallel}, \ldots)$---identical to the adiabatic shape.

Branching between adiabatic and kinetic paths uses Python \texttt{if/else} on \texttt{params.adiabatic\_electrons}, which is a static pytree field resolved at JIT trace time. This ensures zero overhead for the adiabatic path.

\subsection{Field Solver}
The electrostatic potential is obtained algebraically from quasi-neutrality at each RK4 stage.

For adiabatic electrons, the numerator is the velocity-space integral $\sum_{v_\parallel,\mu} Z_i n_i J_0 B \Delta v_\parallel \Delta\mu \cdot \delta f_i$, and the denominator includes both the ion polarization $Z_i^2 n_i (\Gamma_0^i - 1)/T_i$ and the adiabatic electron response $-Z_e n_e / T_e$. The zonal mode ($k_y=0$) receives a flux-surface-averaged correction.

For kinetic electrons, both numerator and denominator sum over all kinetic species. The zonal mode denominator is set to unity (no flux-surface correction needed). Electron FLR effects are retained: $\Gamma_0^e \approx 0.99$--$1.0$ depending on $k_\perp$ (small but nonzero at high $k_\perp$ due to $m_e/m_i \approx 1/3600$).

\subsection{JAX Optimization}
The solver is designed for end-to-end XLA compilation via \texttt{jax.lax.scan}. The precomputed coefficient dictionary is created once outside the scan and captured as a closure variable, ensuring that FFT grid sizes (which must be concrete at trace time) remain available as Python integers rather than traced values.

\texttt{GKParams} uses a custom pytree registration where boolean and string fields (\texttt{non\_linear}, \texttt{adiabatic\_electrons}, \texttt{finit}) are stored as auxiliary data rather than leaves, enabling Python-level control flow branching without tracing issues.

Transport diagnostics (spectra, fluxes) are computed only at block boundaries via \texttt{get\_integrals}, not at every RK4 stage, minimizing overhead. The Parseval factor $P_{k_y}$ and parallel integration weight $\Delta s$ are applied to spectral diagnostics for correct normalization against GKW reference data.

\section{Optimizing JAX with CUDA kernels}
\label{app:cuda_kernels}

As the complexity of the \texttt{gyaradax} right-hand side (RHS) scales, the memory access patterns of the fused finite-difference stencils and the dealiased Fast Fourier Transforms (FFTs) become the primary performance bottlenecks. While XLA provides excellent fusion for element-wise operations, certain patterns in \texttt{gyaradax}---such as high-rank gather-and-dot operations for streaming and mirroring terms---often result in suboptimal memory layouts or redundant global memory accesses.

Such limitations of XLA, particularly for codebases that do not heavily rely on element-wise operations or dense linear algebra, have been documented in the literature \cite{jakob2022dr}. Following approaches outlined in recent work on LLM-driven kernel generation and optimization \cite{ouyang2025kernelbench, zhu2026qimeng, wei2025astra}, we establish automated CUDA kernel optimization loops driven by coding agents.

\textbf{Component-wise solver benchmarks:} First, we establish baselines for the high-level components of the \texttt{gyaradax} solver (see \cref{sec:functions}). We measure execution speed, high bandwidth memory (HBM) utilization, arithmetic intensity, and numerical accuracy (measured via the $L_2$ error relative to the initial, numerically validated JAX implementation).

\textbf{XLA FFI setup:} Paralleling the JAX implementation, we initialize an empty C++ shared library with a functioning XLA Foreign Function Interface (FFI) for CUDA kernels, adhering to the official \href{https://openxla.org/xla/custom_call}{XLA} and \href{https://docs.jax.dev/en/latest/ffi.html}{JAX} documentation. We also provide a CMake configuration to build and compile the CUDA kernels with the appropriate compilers and shared libraries corresponding to our development JAX version.

\textbf{Generating optimization proposals:} Next, we prompt deep reasoning models (Gemini 3.1 Pro, Claude 4.6 Opus) to propose optimizations. To generate highly structured, iterable proposals, we instruct the models to categorize their proposals into three tiers: (1) JAX-level optimizations, (2) CUDA-level optimizations, and (3) algorithmic optimizations. For the CUDA-level, the models typically decompose their proposals into several progressive optimization steps.

\textbf{Testing proposals:} Following the generation phase, we deploy fast coding models (Gemini 3 Flash, Claude 4.5 Sonnet) to implement the Tier 1 and Tier 2 proposals. Empirically, the JAX-level optimizations yielded marginal improvements; we attribute this to the highly optimized nature of the XLA compiler's existing passes and the high quality of the initial JAX code generated by the agents out of the box.

Conversely, the iterative CUDA kernel optimization approach proved highly successful, yielding significant speedups across various solver components. In the following subsections, we outline the specific operation types for which XLA struggles to compile efficient kernels, and detail how we replaced these with LLM-generated custom CUDA kernels.

\subsection{Fused Linear RHS Kernel}
The linear right-hand side evaluation combines a nine-point parallel stencil, a five-point velocity-space stencil, and eight elementwise physics terms (drift advection, trapping, hyper-diffusion, equilibrium drive, and field-drive coupling) into a single expression. XLA's compiler already fuses most of this computation into one monolithic kernel, but inspection of the HLO intermediate representation revealed two structural inefficiencies it cannot resolve. 

First, XLA implements the velocity-space stencil through general-purpose gather operations with runtime boundary predicates, generating five full-array index computations and conditional selects where a simple stride-based access with compile-time boundary checks suffices. Second, XLA passes all coefficient arrays at their natural declaration shapes—some lacking the magnetic moment dimension, others lacking radial or toroidal wavenumber axes—and broadcasts them element-wise within the fused kernel, resulting in redundant address arithmetic and suboptimal register reuse across the velocity dimension. 

Our fused CUDA kernel addresses both limitations in a single pass: the velocity stencil uses direct strided addressing with compile-time boundary constants, eliminating the gather and predicate machinery entirely. All coefficient arrays are stored in their minimal shapes, with index reconstruction computed from the thread identity. Velocity-axis tiling further amortizes the parallel stencil map lookups across eight consecutive velocity indices sharing the same magnetic moment, converting what XLA executes as independent per-element gathers into sequential streaming reads. The resulting kernel outperforms XLA's auto-fused baseline on the full linear RK4 step while maintaining numerical parity up to machine precision.

\subsection{Nonlinear RHS: Fused FFTs }
\label{sec:nonlinear_fft}
 
The nonlinear $\mathbf{E}\times\mathbf{B}$ advection term is evaluated pseudospectrally via a Poisson bracket that requires four inverse FFTs (to obtain real-space gradients $\partial_x\phi$, $\partial_y\phi$, $\partial_x f$, $\partial_y f$), a pointwise bracket multiplication, and one forward FFT. XLA's kernel fusion has a fundamental limitation here: the FFTs are dispatched as opaque calls to cuFFT, which acts as a hard fusion barrier. Therefore, the spectral derivative multiplications preceding each inverse FFT and the bracket multiplication following them must be compiled as separate kernels
 
\paragraph{LTO callback fusion.}
We replace this pipeline with three fused cuFFT kernel launches using Link-Time Optimization (LTO) device callbacks ~\cite{cufft_lto}. LTO callbacks are compiled to device-relocatable code (\texttt{-dlto}) and linked into the cuFFT plan at plan-creation time via \texttt{cufftXtSetJITCallback}. This allows arbitrary user logic to execute \emph{inside} cuFFT's butterfly passes, at the exact point where each spectral element is loaded from or stored to global memory.
 
\begin{figure}[htbp]
    \centering
    \begin{subfigure}{0.48\textwidth}
        \centering
        \includegraphics[width=\linewidth]{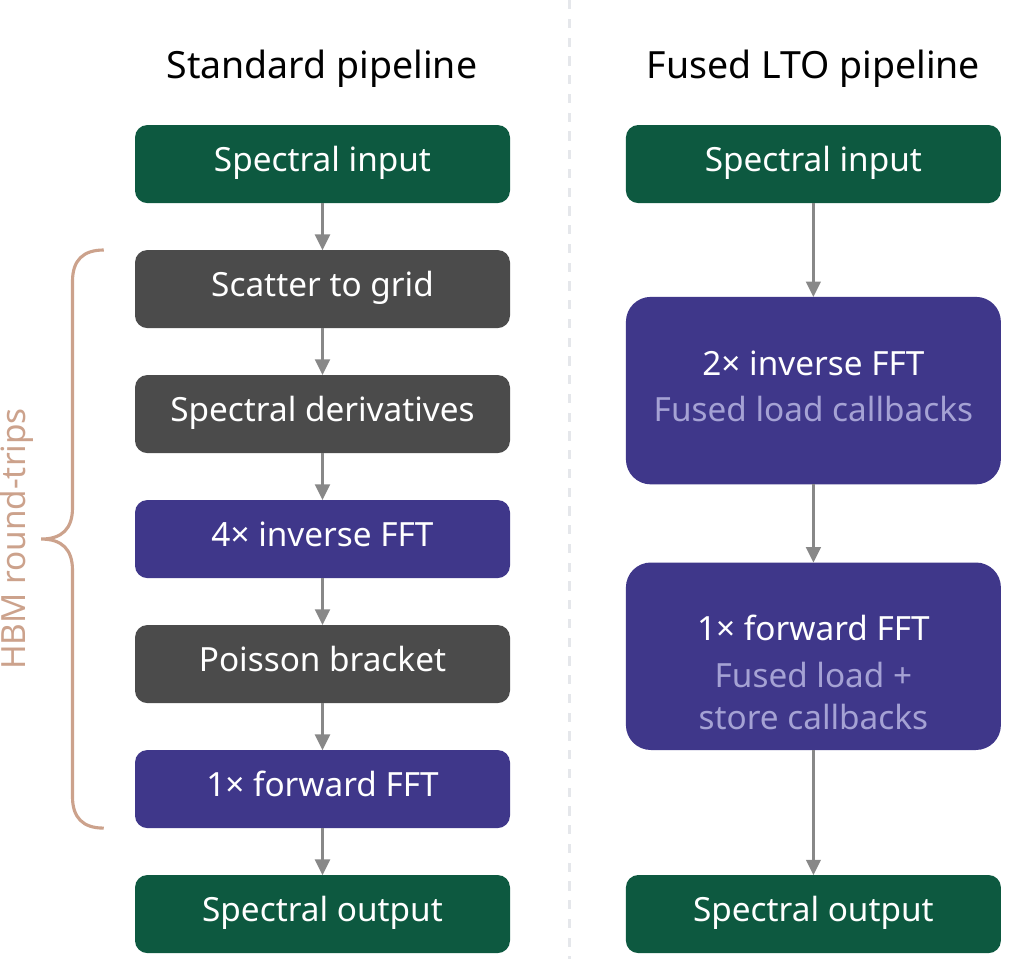}
        \caption{Fused LTO pipeline}
        \label{fig:cufft_lto}
    \end{subfigure}\hfill
    \begin{subfigure}{0.48\textwidth}
        \centering
        \includegraphics[width=\linewidth]{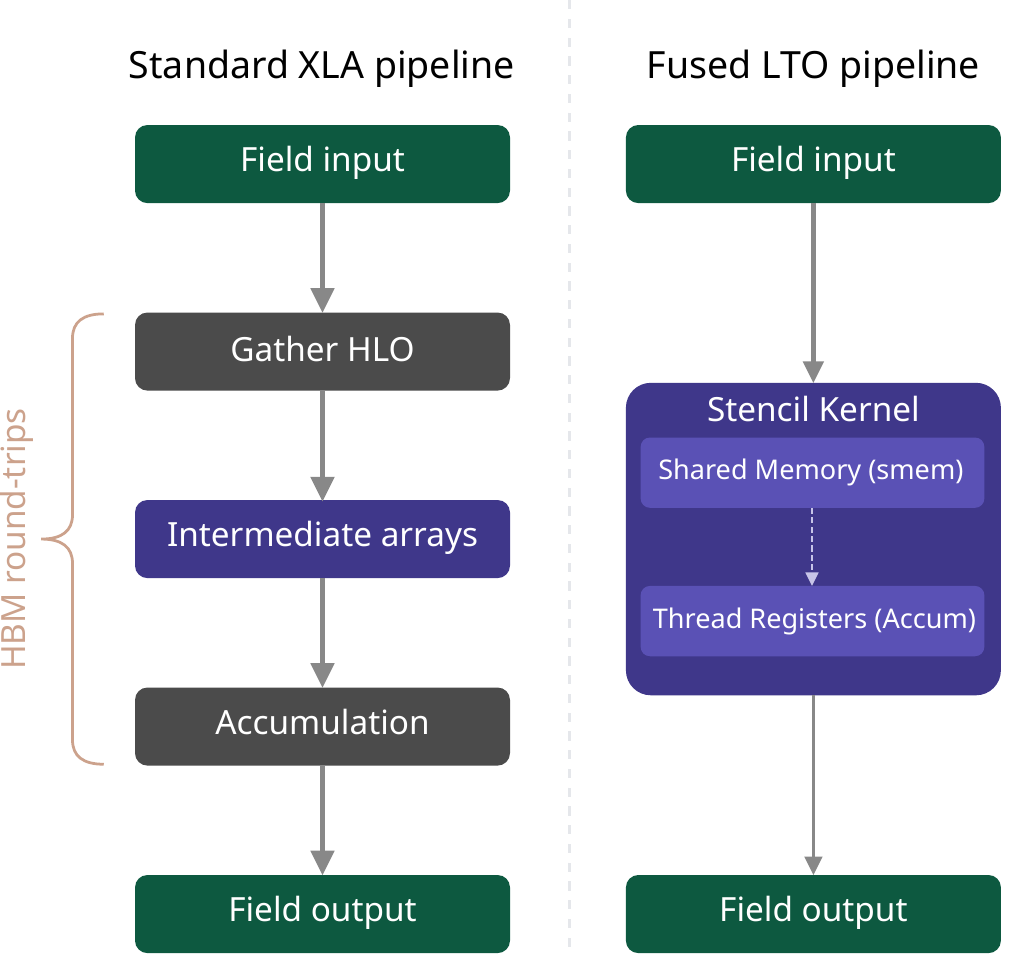}
        \caption{Stencil CUDA fusion}
        \label{fig:stencil_fusion}
    \end{subfigure}
    \caption{Overview of our memory optimization strategies.}
    \label{fig:optimizations}
\end{figure}

We exploit this mechanism at three points in the pipeline (see \cref{fig:cufft_lto}):
\begin{enumerate}
    \item \textbf{Inverse FFT load callback:} Fuses the scatter-to-dense-grid, spectral derivative multiplication ($ik_x$, $ik_y$), gyro-averaging (Bessel $J_0$ multiplication), and amplitude scaling into a single register-only computation. Each packed spectral element is read from HBM exactly once; the callback returns the fully processed value directly to cuFFT's butterfly registers.
    \item \textbf{Forward FFT load callback:} Fuses the Poisson bracket computation. Instead of materializing four real-space gradient arrays and computing $\partial_y\phi\,\partial_x f - \partial_x\phi\,\partial_y f$ in a separate kernel, the callback reads the gradient arrays and computes the bracket in-flight, returning the result to the forward FFT without an intermediate HBM round-trip.
    \item \textbf{Forward FFT store callback:} Writes directly to the packed output spectrum, skipping the 59\% of dealiased modes that would otherwise be written and immediately discarded.
\end{enumerate}
 
\subsection{Two-for-One Spectral Packing}

To further reduce the cuFFT overhead, the reasoning model proposed exploiting the linearity of the discrete Fourier transform (DFT). If two signals $A$ and $B$ are real-valued in physical space, their spectral representations are Hermitian-symmetric. The agent's optimization combines them into a single complex signal $Z_k = A_k + i B_k$, which can be inverse-transformed via a single complex-to-complex (C2C) FFT to recover $A = \mathrm{Re}(\mathrm{IFFT}(Z))$ and $B = \mathrm{Im}(\mathrm{IFFT}(Z))$. Specifically, we pack both spatial derivatives of the \emph{same} physical field into each C2C transform:

$$Z_k^{(\phi)} = \frac{ik_x \hat\phi_k}{\alpha_\phi} + i\,\frac{ik_y \hat\phi_k}{\beta_\phi}, \quad Z_k^{(f)} = \frac{ik_x \hat f_k}{\alpha_f} + i\,\frac{ik_y \hat f_k}{\beta_f}$$

where $\alpha$ and $\beta$ are dynamic scaling factors computed from the spectral magnitudes of each field. This strategy successfully halves the inverse FFT count from four to two. The scaling factors ensure both channels occupy comparable dynamic ranges, preventing floating-point leakage. Because both derivatives originate from the same field, the ratio $\alpha/\beta$ is inherently bounded by the grid anisotropy $\max|k_x|/\max|k_y|$, an $\mathcal{O}(1)$ property of the grid geometry.

\paragraph{Hermitian symmetrization.}
The C2C transform operates on the full two-dimensional spectrum rather than the half-spectrum used by the standard real-to-complex path. The load callback synthesizes the conjugate half via Hermitian extension. For this packing to produce mathematically exact results, the spectrum of each derivative must be perfectly Hermitian-symmetric; otherwise, imaginary leakage crosses into the other channel. 

During the implementation of this agent-proposed kernel, the automated evaluation harness consistently flagged numerical divergence. Investigating this strict validation failure led to the discovery that the gyro-averaged potential $\hat\phi_k$ exhibits a small but non-negligible symmetry defect (up to $14\%$ relative error) at the zeroth wavenumber, introduced by the Bessel function multiplication during gyro-averaging. 

While entirely invisible to the baseline JAX real-to-complex inverse FFT---which inherently discards the imaginary output by construction---this asymmetry produced a parasitic imaginary component in the C2C output that corrupted the packed channels. To resolve this and satisfy the evaluation harness, an explicit Hermitian symmetrization step was added to the load callback: for each mirror pair $(k_x, -k_x)$, both values are dynamically replaced with their Hermitian average. This algorithmic correction, directly prompted by the agentic verification loop, eliminated the channel leakage entirely and reduced the relative error from $\mathcal{O}(10^{-4})$ to $\mathcal{O}(10^{-14})$ in FP64 and from $\mathcal{O}(10^{-2})$ to $\mathcal{O}(10^{-6})$ in FP32.

For the mixed precision configuration, this results in an accumulated error  of $\mathcal{O}(10^{-10})$ for an entire RK4 step, maintaining overall solver stability.

\section{Comparison with GKW Reference Code}
\label{app:gkw_comparison}

While \texttt{gyaradax} aligns with the local flux-tube physics of the GKW framework, several advanced features of the Fortran reference code are identified as future enhancements.

\textbf{Physics.}
GKW supports full electromagnetic dynamics, including the parallel vector potential ($A_\parallel$) and perpendicular magnetic perturbations ($B_\parallel$); \texttt{gyaradax} is currently limited to the electrostatic limit ($\beta = 0$). GKW features linearized collision operators including pitch-angle scattering and energy diffusion with conservation terms, whereas \texttt{gyaradax} is collisionless. Rotation physics---Coriolis drifts and centrifugal forces arising in rotating frames (Term~VI in the GKW numbering)---is implemented in GKW but omitted here. Background Krook sources and buffer regions for global simulations are also absent.

\textbf{Numerical methods.}
\texttt{gyaradax} uses an explicit RK4 scheme exclusively. GKW additionally supports implicit and semi-implicit integrators for stiff parallel streaming, as well as Runge-Kutta-Chebyshev (RKC) schemes for diffusion-dominated regimes. GKW also supports global simulations with non-uniform background gradients and Dirichlet boundary conditions; \texttt{gyaradax} is restricted to the local flux-tube approximation.

\textbf{HPC and scalability.}
GKW features highly optimized MPI-based domain decomposition across species, field-line, and velocity dimensions. While JAX provides native array sharding via \texttt{jax.sharding}, distributing the 6D kinetic grid across multiple GPUs remains a planned enhancement. \texttt{gyaradax} uses mixed precision for the nonlinear FFTs (FP32) while keeping linear terms in FP64, which could be further extended to cover additional hot paths.

\section{Analytical Benchmark Details}
\label{app:analytical_details}

\subsection{Rosenbluth-Hinton Zonal Flow Test}

The Rosenbluth-Hinton test \cite{Rosenbluth} initialises a radial ($k_\psi \neq 0$, $k_\zeta = 0$) density perturbation that excites a geodesic acoustic mode (GAM). The mode oscillates at the GAM frequency and damps via collisionless Landau damping, leaving a residual zonal potential given analytically by \cite{xiao2006plasma}:
\begin{equation}
\frac{\phi(\infty)}{\phi(0)} = \frac{1}{1 + q^2 \Theta / \varepsilon^2}, \qquad \Theta = 1.6\,\varepsilon^{3/2} + 0.5\,\varepsilon^2 + 0.36\,\varepsilon^{5/2}.
\end{equation}

We use the GKW benchmark parameters from \texttt{gkw\_ref/benchmarks/zonal\_flow/zonal01}: $q{=}1.3$, $\hat{s}{=}0.1592$, $\varepsilon{=}0.05$, s-alpha geometry, $N_s{=}128$, $N_{v\parallel}{=}128$, $N_\mu{=}16$, $k_\psi\rho_s{\approx}0.025$ (\texttt{krhomax=0.025}, \texttt{ikxspace=1}), $\nu_\parallel{=}0.01$ (\texttt{disp\_par}), $\mathrm{d}t{=}0.01$, \texttt{finit='zonal'}, \texttt{amp\_init=$10^{-4}$}. The parallel dissipation damps velocity-space filamentation, enabling convergence to the analytical residual. The simulation runs for $t{=}100$ ($R/v_\text{th}$). At $t{>}80$, the residual converges to 0.0711, matching the Xiao-Catto prediction to 0.1\%.

The $\varepsilon$-scan (\cref{fig:rh}b) uses the same parameters with $\varepsilon \in \{0.05, 0.1, 0.15, 0.2, 0.25\}$, each run to $t{=}100$.

\subsection{Cyclone Base Case Linear ITG}

The Cyclone Base Case \cite{cbc} is the standard linear benchmark: $q{=}1.4$, $\hat{s}{=}0.78$, $\varepsilon{=}0.19$, $R/L_T{=}6.9$, $R/L_n{=}2.2$, $T_e/T_i{=}1$, electrostatic, adiabatic electrons, s-alpha geometry.

We use the GKW benchmark parameters from \texttt{gkw\_ref/benchmarks/cyclone/linear}: $N_s{=}160$, $N_{v\parallel}{=}64$, $N_\mu{=}16$, $n_\text{period}{=}5$, $\nu_\parallel{=}1.0$, $\mathrm{d}t{=}0.003$, \texttt{naverage=100}. The high parallel resolution ($N_s{=}160$ across 5 poloidal periods) is essential for resolving Landau damping; at lower resolution (e.g.\ $N_s{=}16$), growth rates are systematically overestimated due to insufficient velocity-space phase mixing. Note that \texttt{naverage} $\geq 10$ is required for correct growth rate measurement, as the ITG mode's finite real frequency causes the amplitude to oscillate within a single timestep.

The $k_\theta\rho_s$ scan (\cref{fig:cbc}a) covers $[0.1, 0.8]$ at fixed $R/L_T{=}6.9$. The $R/L_T$ scan (\cref{fig:cbc}b) covers $\{6.9, 8.28, 10.35, 12.44, 15.18\}$ at fixed $k_\theta\rho_s{=}0.5$. Each run evolves 300 \texttt{naverage} windows. Growth rates match the GKW reference (run with identical parameters) to within 1\%.

\newpage
\section{Validation Overview}
\label{app:validation_overview}

\begin{figure}[H]
    \centering
    \includegraphics[height=0.9\textheight]{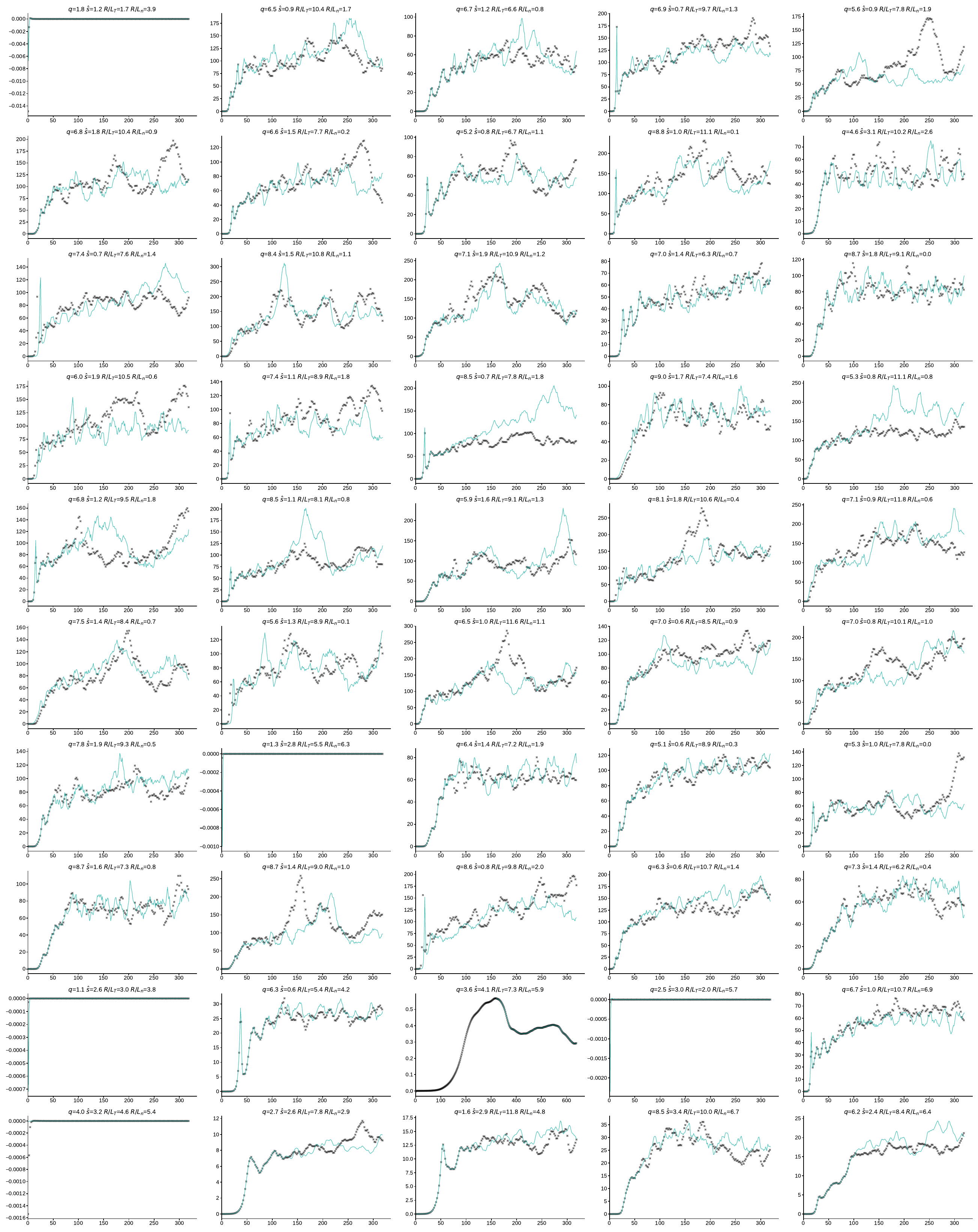}
    \caption{Heat flux time traces for all adiabatic validation configurations. Each panel shows \texttt{gyaradax} (cyan) versus GKW (black markers) for a distinct combination of $q$, $\hat{s}$, $R/L_T$, and $R/L_n$.}
    \label{fig:appx_fluxes}
\end{figure}

\begin{figure}[H]
    \centering
    \includegraphics[height=0.95\textheight]{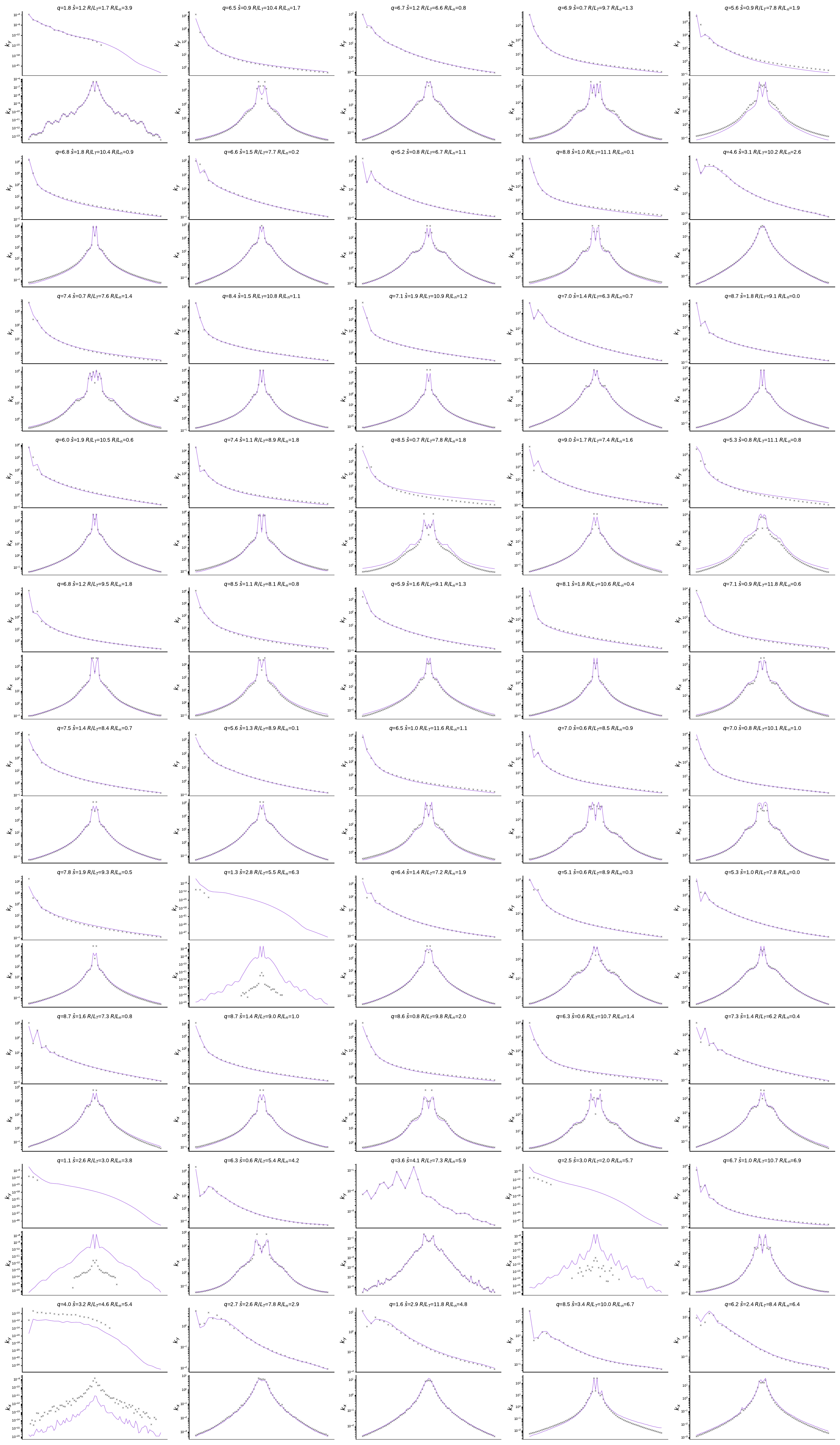}
    \caption{Time-averaged $k_y$ (top) and $k_x$ (bottom) spectral profiles for all adiabatic validation configurations. \texttt{gyaradax} (purple lines) versus GKW (black markers).}
    \label{fig:appx_spectra}
\end{figure}

\newpage
\section{\texttt{PROMPT.md}}
\label{app:prompt}
\begin{promptbox}[]
\scriptsize\sffamily
\textbf{\normalsize PROJECT MANDATES: JAX Reimplementation of GKW}

\vspace{0.5em}
\textbf{Documentation \& Note-Taking}
\begin{itemize}[leftmargin=*, noitemsep, topsep=0pt]
    \item \textbf{Thoroughness is Mandatory:} You must be extremely thorough in your note-taking. Do not hesitate to produce long, detailed files.
    \item \textbf{Granular Detail:} Document everything required for reimplementation and more: specific subroutine logic, module dependencies, variable mappings, numerical constants, and edge-case handling.
    \item \textbf{Foundational Reference:} Notes in \texttt{GKW.md} (and similar files) serve as the foundational technical specification for the JAX port.
\end{itemize}

\vspace{0.5em}
\textbf{CONTEXT FILES PROVIDED:}
\begin{itemize}[leftmargin=*, noitemsep, topsep=0pt]
    \item \texttt{@jax\_integrals.py} (Contains the implemented JAX flux integrals).
    \item \texttt{@utils.py}, \texttt{@jax\_geometry.py} (Data, potential/phi, and geometry loading helpers).
    \item \texttt{@test\_jax\_integral.py}
    \item \texttt{@test\_linear.py}, \texttt{@test\_nonlinear.py} (Empirically verify content against GKW reference trajectories).
\end{itemize}

\vspace{0.5em}
\textbf{ROLE:} Act as an autonomous, expert Scientific Computing Engineer. Your objective is to translate the GKW Fortran code into JAX. Describe what you plan to do at the main intermediate steps. \textbf{Crucially, stop and yield to the user for feedback at the end of each numbered execution phase before proceeding to the next.}

\vspace{0.5em}
\textbf{SCOPE \& CONSTRAINTS (STRICT):}
\begin{itemize}[leftmargin=*, noitemsep, topsep=0pt]
    \item \textbf{Physics:} Adiabatic electron case ONLY. Simplify all equations to reflect this.
    \item \textbf{Grid Resolution:} (vpar, mu, s, x, ky) = (32, 8, 16, 85, 32)
    \item \textbf{Time step:} dt = 0.01
    \item \textbf{Framework:} JAX. All functions must be pure, differentiable, and \texttt{jit}-compatible.
    \item \textbf{Precision:} ALWAYS use float64 (fp64). JAX must be configured to use 64-bit precision.
    \item \textbf{Integrity:} STRICTLY NO normalization sweeps to force test passes. NO cheating or hacking the outputs to match expected results. NO hardcoded constants—derive everything mathematically from the source code and physical constants.
    \item \textbf{Environment:} STRICTLY use the environment under \texttt{<PATH\_TO\_ENV>} for everything. STRICTLY use GPU:0 for everything.
\end{itemize}

\vspace{0.5em}
\textbf{ENVIRONMENT:} \\
Use \texttt{@utils.py} for data loading and potential (phi) calculations. Do NOT reimplement these helpers. \\
\textbf{CRITICAL:} The flux integrals have already been implemented in JAX within \texttt{@jax\_integrals.py}. Your first coding task is to verify that they function correctly before proceeding to the main solver.

\vspace{0.5em}
\textbf{REQUIRED INTERFACE:} \\
You MUST expose the following purely functional interface for the core solver: \\
\texttt{next\_df, (phi, fluxes) = gksolve(prev\_df, ...)}

\vspace{0.5em}
\textbf{EXECUTION PLAN (FOLLOW EXACTLY IN ORDER):} \\
Do not write the final solver code immediately. You must execute the following steps sequentially. Use your toolset to explore directories, read files, write code, and run tests. Output and save your notes and progress to markdown artifacts for each step before pausing.

\begin{enumerate}[leftmargin=*, noitemsep, topsep=0pt]
    \item \textbf{INITIAL CONTEXT INGESTION:} Before starting any code translation or detailed planning, explore the \texttt{gkw\_ref/src} (Fortran code) and \texttt{gkw\_ref/manual} (LaTeX files) directories. You MUST fully load the following specific files into your context, and explore for any other useful ones:
        \begin{itemize}[leftmargin=*, noitemsep, topsep=0pt]
            \item \texttt{gkw\_ref/src/gkw.f90} (Main loop over large time steps, normalisation).
            \item \texttt{gkw\_ref/src/exp\_integration.F90} (Explicit integration schemes like rk4, RHS assembly order).
            \item \texttt{gkw\_ref/src/linear\_terms.f90} (Linear operators, field coupling matrices, poisson splits).
            \item \texttt{gkw\_ref/src/fields.F90} (Field solve application, zonal adiabatic correction).
            \item \texttt{gkw\_ref/src/components.f90} (Adiabatic electrons flag, species setup).
        \end{itemize}
        Individuate any additional files that are relevant to time integration and the adiabatic electron formulation. \textbf{Stop and wait for the user to approve the final identified file list.}
    
    \item \textbf{VERIFY JAX FLUX INTEGRALS:} 
    \begin{itemize}[leftmargin=*, noitemsep, topsep=0pt]
        \item Review the existing JAX integrals in \texttt{@jax\_integrals.py}.
        \item Verify that the geometry loading functions from \texttt{@jax\_geometry.py} work seamlessly with this JAX implementation.
        \item Run and adapt \texttt{@test\_jax\_integral.py} if necessary to confirm the JAX integrals pass all tests.
        \item TAKE NOTES on any specific broadcasting, memory layout, or numerical precision details observed during verification. \textbf{Stop and wait for user approval.}
    \end{itemize}
    
    \item \textbf{EXPLORATION, ANALYSIS \& PLANNING (NOTE-TAKING):}
    \begin{itemize}[leftmargin=*, noitemsep, topsep=0pt]
        \item \textbf{Source Code \& Theory:} Actively analyze the Fortran source code and LaTeX manual files you loaded in Step 1. Focus strictly on time integration and the adiabatic electron formulation.
        \item \textbf{Reference Data:} Ingest and explore the reference trajectory at \texttt{<PATH\_TO\_DATA>/iteration\_13} to understand data structures, empirical array shapes, and physical scales.
        \item \textbf{Synthesize \& Plan:} Based on the Fortran code, the LaTeX manual, and the reference data, write out a high-level plan for the core JAX architecture. Identify exactly which Fortran modules and subroutines map to the \texttt{gksolve} update step, detail the exact mathematical update equations, and map out the variables.
        \item \textbf{Artifact Generation:} Save all of these findings into a detailed \texttt{GKW.md} file. \textbf{Stop and wait for user approval.}
    \end{itemize}
    
    \item \textbf{ESTABLISH CORE TESTS (TDD):} Write the test suite for the core simulator before implementing it. These tests must run independently.
    \begin{itemize}[leftmargin=*, noitemsep, topsep=0pt]
        \item \textbf{Mandatory:} Write a validation test that checks calculated growth rates against the \texttt{growth\_rate\_all\_modes} file.
        \item \textbf{Mandatory:} Write unit tests verifying array shapes (based on your notes from Step 3) and basic conservation properties. \textbf{Stop and wait for user approval.}
    \end{itemize}
    
    \item \textbf{IMPLEMENT LINEAR \texttt{gksolve}:} Write the \texttt{gksolve} function for the LINEAR case, based on the notes you created (Terms I, II, IV, V, VII, VIII, Diffusion). Proceed to this step ONLY after all prior tests are passing. \textbf{Stop and wait for user approval.}
    
    \item \textbf{IMPLEMENT NONLINEAR \texttt{gksolve}:} Finalize \texttt{gksolve} for the NONLINEAR case by extending the LINEAR version with Term III. Proceed to this step ONLY after all prior tests are passing, ESPECIALLY \texttt{@test\_linear.py}. TO CONCLUDE, make sure that both \texttt{@test\_linear.py} and \texttt{@test\_nonlinear.py} PASS. \textbf{Report the final test results.}
\end{enumerate}
\end{promptbox}

\end{document}